\documentclass[12pt,oneside,a4paper]{article}
\usepackage{amsfonts,amssymb,latexsym}
\usepackage{graphicx}
\usepackage{ae}
\usepackage{geometry}
\usepackage{amsmath,bbm}

\numberwithin{equation}{section}

\newcommand{\tpi}{2\pi\I}
\newcommand{\Iouter}{I_{>}}
\newcommand{\Ismall}{I_{<}}

\newcommand{\Fmatrix}{\mathbb{A}}

\newcommand{\Snorm}[2]{\|#1 \|_{\mathcal{S},#2}}
\newcommand{\Sdnorm}[3]{\|#1 \|_{\mathcal{S}_{#2},#3}}
\newcommand{\Tt}{\ensuremath{\mathbb{T}^3}}
\newcommand{\K}{\ensuremath{\mathbb{K}}}
\newcommand{\Sbb}{\ensuremath{\mathbb{S}}}
\newcommand{\sabs}[1]{\langle #1\rangle}

\newcommand{\veps}{\varepsilon}
\newcommand{\vep}{\varepsilon}

\newcommand{\mean}[1]{\langle #1\rangle}
\newcommand{\muH}{\mu^{\mathrm H}}
\newcommand{\wlim}{\rightharpoonup}
\newcommand{\wslim}{\overset{*}{\rightharpoonup}}
\newcommand{\cutoffone}{f}
\newcommand{\cutoff}{\varphi}
\newcommand{\ccutoff}{\tilde{\varphi}}
\newcommand{\cutoffrac}[2]{\cutoff\left(\tfrac{#1}{#2}\right)}
\newcommand{\ccutoffrac}[2]{\ccutoff\left(\tfrac{#1}{#2}\right)}

\def\D{\mathrm{d}}
\def\I{\mathrm{i}}
\def\E{\mathrm{e}}

\def\wh{\hat}

\def\dd{\mathrm{d}}

\def\R{\ensuremath{\mathbb{R}}}
\def\T{\ensuremath{\mathbb{T}^3}}

\def\schwartz{\ensuremath\mathcal{S}}
\def\fourier{\ensuremath\mathcal{F}}

\newcommand{\frechet}{Fr\'{e}chet}

\newcommand{\Wcont}{\ensuremath{W_{\text{cont}}}}

\def\Wep{\ensuremath{W^{\veps}}}

\def\Z{\ensuremath{\mathbb{Z}}}
\def\C{\ensuremath{\mathbb{C}}}

\newcommand{\norm}[1]{\Vert #1\Vert}
\renewcommand{\mod}[1]{| #1|}

\newtheorem{thm}{Theorem}[section]
\newtheorem{assump}[thm]{Assumption}
\newtheorem{lemma}[thm]{Lemma}
\newtheorem{prop}[thm]{Proposition}
\newtheorem{defn}[thm]{Definition}
\newtheorem{remark}[thm]{Remark}
\newtheorem{cor}[thm]{Corollary}

\newenvironment{proof}{\begin{trivlist}\item[]{\em Proof:}\/}{%
\hfill\mbox{$\Box$}\end{trivlist}}

\begin{document}

\newcommand{\emailjani}{jlukkari@ma.tum.de}
\newcommand{\addressjani}{Zentrum Mathematik, Technische Universit\"at
  M\"unchen, Boltzmannstr. 3, D-85747 Garching, Germany}
\newcommand{\addressflorian}{Warwick University, Mathematics Institute,
  Coventry, CV4 7AL, UK}
\newcommand{\addressstefan}{Mathematisches Institut, Auf der Morgenstelle
  10, 72076 T\"{u}bingen, Germany}

\title{Energy transport by acoustic modes of harmonic lattices}

\author{Lisa Harris\thanks{\addressflorian},
Jani Lukkarinen\thanks{\addressjani},
Stefan Teufel\thanks{\addressstefan},
and Florian Theil\protect\footnotemark[1]}

% % For amsart
% \author{Lisa Harris}
% \address{\addressflorian}

% \author{Jani Lukkarinen}
% \address{\addressjani}

% \author{Stefan Teufel}
% \address{\addressstefan}

% \author{Florian Theil}

%Stefan Teufel, and Florian Theil}

% \begin{center}
% Lisa Harris\footnote[1]{Warwick University, Mathematics Institute,
% Coventry, CV4 7AL, UK}, Jani Lukkarinen\footnote[2]{TU M\"{u}nchen,
% Zentrum Mathematik, Boltzmannstr. 3, 85747 Garching, Germany}, Stefan
% Teufel\footnote[3]{Mathematisches Institut, Auf der Morgenstelle 10,
% 72076 T\"{u}bingen, Germany} and Florian
% Theil\footnotemark[1]\end{center}

\maketitle

\begin{abstract}
We study the large scale evolution of a scalar lattice excitation
$u$ which satisfies a discrete wave-equation in three
dimensions, $\ddot u_t(\gamma) = -\sum_{\gamma'}
\alpha(\gamma-\gamma') u_t(\gamma')$, where $\gamma,\gamma'\in
\Z^3$ are lattice sites. We assume that the dispersion relation
$\omega$ associated to the elastic coupling constants
$\alpha(\gamma-\gamma')$ is acoustic, i.e., it has a singularity of
the type $|k|$ near the vanishing wave vector, $k= 0$.

To derive equations that describe the macroscopic energy transport
we introduce the Wigner transform and change variables so that the
spatial and temporal scales are of the order of $\veps$. In the
continuum limit, which is achieved by sending the parameter
$\veps$ to $0$, the Wigner transform disintegrates into three
different limit objects: the transform of the weak limit, the
H-measure and the Wigner-measure. We demonstrate that these three
limit objects satisfy a set of decoupled transport equations: a
wave-equation for the weak limit of the rescaled initial data, a
dispersive transport equation for the regular limiting Wigner
measure, and a
%non-dispersive
geometric optics transport equation  for the H-measure
limit of the initial data concentrating to $k=0$.

A simple consequence of our result is the complete
characterization of energy transport in harmonic lattices with
acoustic dispersion relations.
\end{abstract}

\section{Introduction.}

The energy transport by atomistic oscillations in crystalline
solids is a central question in solid state physics. To the first
order approximation, the oscillations can be described by a
discrete wave equation 
$$\ddot u_t(\gamma) = -\sum_{\gamma'}
\alpha(\gamma-\gamma') u_t(\gamma')$$
where $u(\gamma) \in \R$ is composed out of the displacements
of the crystal atoms from their equilibrium position, as will be discussed in
Sec.~\ref{sec:physics}.   
To analyze physically relevant properties of the crystal, such as its
thermal conductivity, we first need to understand how energy is transported
within the crystal via purely harmonic vibrations. 
Such transport properties are determined by the dispersion relation $\omega$
of the crystal, here $\omega(k)=\sqrt{\hat{\alpha}(k)}$, the ``hat''
denoting a discrete Fourier transform.
If $\omega$ is not smooth, then depending on the wavelength
different types of continuum
energy transport equations can arise.
We follow the basic
ideas of Luc Tartar who developed in the 1980s a mathematical
framework that can be used to analyze the weak limits of certain
nonlinear quantities, the energy density being one of them.

Our main interest is to characterize the macroscopic evolution of
the energy density. 
The starting point of our mathematical analysis is the {\em Wigner
transform} which can be interpreted as a ``wavenumber resolved'' energy
density.  Let us leave the details for Sec.~\ref{sec:energy}, and only
summarize the main findings here.
The Wigner transform $\Wep=\Wep[\psi]$ of the
field $\psi\in \ell_2(\Z^3)$ corresponding to a given normal mode
allows defining the corresponding energy density, $e^\vep =
e^\vep[\psi]$, by the formula
\begin{align}\label{eq:edWig1}
e^\vep\!\left(x\right)=\int_{\T}\dd k \, \Wep(x,k),
\end{align}
where $\vep>0$ denotes the ``lattice spacing'' and $x\in \R^3$ is
a variable which interpolates between the points on the scaled
lattice $\vep \Z^3$. Here $\T$ denotes the $3$-torus, and we
identify $\T=\R^3/\Z^3$. Thanks to the equality
$\int_{\R^3} \dd q \, \int_{\T}\dd k \, \Wep(x,k)= H(u_{t=0}, \dot
u_{t=0})$ we are able to identify $e^\vep(x)$ as the energy
density at position $x \in \R^d$. 

A limit of a sequence $(\Wep[\psi^\vep])$, where $\veps$ tends
to $0$, is in general given by a 
non-negative Radon-measure $\mu \in M_+$.  For such limit measures to form
a sensible approximation for the original dynamical system, the convergence
property needs to be 
retained in the time-evolution.  In addition, for such a description to be
useful, $\mu_t$ should also satisfy an autonomous
evolution equation.  However, this is typically not possible if the initial
measure concentrates to the singular set of $\omega$, i.e., to the points
where $\omega$ is not smooth.  Here we will augment the above Wigner
transform scheme to encompass the most common type of singularity 
encountered in solid state physics: the case when $\omega$ behaves like
$|k|$ near $k=0$.  Such modes will occur in general within crystal models
with short range interactions, and they are particularly important as they
are responsible for sound propagation in the crystal.  
With some effort, it is likely that our results could be extended to cover
any dispersion relation for which the singular set consists of isolated
points.   However, we will not explicitly spell out the general result here.

Our result can be seen as a generalization of the analysis in
\cite{mielke05} where it is shown that $\mu_t$ can be computed
from $\mu_0$ by solving a dispersive linear transport equation
provided that $\mu$ has no concentrations at wavenumbers $k$ where
the dispersion relation $\omega$ is not $C^1$.

If the dispersion relation is not almost everywhere $C^1$ (with
respect to the initial Wigner-measure), we have to resolve finer details
of the asymptotic behavior of the sequence of initial conditions.
More precisely, we show that if the sequence of initial
excitations is bounded and tight in $\ell_2(\vep \Z^3)$ then for
all $t\in \R$ there are two measures, a Wigner-measure $\mu_t$ on
$\R^3\times \T_*$ and an H-measure $\muH_t$ on $\R^3\times S^2$,
and an $L^2$-function $\phi_t$ such that $\Wep[\psi^\vep(t/\vep)]$
converges along a subsequence to $(\mu_t,\muH_t,\phi_t)$ in a
certain weak sense (Theorem~\ref{th:mainres}).  The subsequence
can be chosen independently of $t$, and it will only be relevant
for determining the limit of the initial data, that is,
$(\mu_0,\muH_0,\phi_0)$. For all other times $t\in\R$, the
measures $\mu_t, \muH_t$ and the $L^2$-function $\phi_t$ can be
determined using the transport equations
\begin{alignat}{2}
\label{tran1} &\partial_t \mu_t(x,k) +  \tfrac{1}{2\pi} \nabla
\omega(k) \cdot \nabla_{\!x} \mu_t(x,k) =0,
\qquad & k\in \T_*\, ,\\
\label{tran2} &\partial_t \muH_t(x,q)  +  \tfrac{1}{2\pi} \nabla
\omega_0(q)\cdot \nabla_{\!x} \muH_t(x,q)  =0,
&  q\in S^2\, ,\\
\label{tran3} &\partial^2_t\phi = \mathrm{div}(\tfrac{1}{(2\pi)^2}
A_0 \nabla \phi),
\end{alignat}
together with the initial conditions
\begin{align}%\label{eq:}
\left.\mu\right|_{t=0} =\mu_0, \quad \left.\muH\right|_{t=0}=\muH
_0, \quad \left.\phi\right|_{t=0}=\phi_0, \quad
\partial_t \hat{{\phi}}(t,q)|_{t=0} = -i\omega_0(q) \hat \phi_0(q).
\end{align}
%In all equations $x \in \R^3$ and $t\in\R$.
Here the constant matrix $A_0$ and the function $\omega_0$ are
determined by the Hessian of the square of the dispersion relation
$\omega$ (defined in (\ref{eq:dispersion})) at $k=0$, explicitly
\begin{align}%\label{eq:}
A_0=\frac{1}{2}D^2\omega^2(0),\qquad \omega_0(q)=\sqrt{q\cdot A_0
q}.
\end{align}

Equation (\ref{tran1}) describes the propagation of energy along
the harmonic lattice with the group velocity
$\nabla\omega(k)/(2\pi)$. Equation (\ref{tran3}) is the wave
equation which describes the evolution of macroscopic
fluctuations. Equation (\ref{tran2}) is usually known under the
name ``geometric optics'' and it describes the evolution of
macroscopic fluctuations whose wavelength is much longer than the
lattice spacing $\veps$ and much smaller than 1, the wavelength of
the fluctuations resolved by $\phi$.

It follows easily from our analysis that the sum of the energies
of $\mu$, $\muH$ and $\phi$ is a constant of motion and equals the
limiting value of the total energy of the initial excitations.

Standard results in semi-classical analysis show that the Wigner
measure $\mu$ and $\int_{\R^3}\dd q\, W^1[\phi](x,q)$ are
non-negative. Hence, a key step in our analysis is to demonstrate
that also $\muH$ has a sign. The measure $\muH$ is closely related
to the H-measure which was introduced by L. Tartar~in
\cite{tartar90} and P.~G\'erard in \cite{gerard88}, but in general
it differs from the H-measure. 

The Wigner transform, or
the Wigner function, was originally introduced to study
semi-classical behavior in quantum mechanics but it has been
proven to be a useful tool in studying large scale behavior of
wave equations, as well \cite{ryzhik96,gerard97}. In particular,
the method of calculating continuous, macroscopic energy by
finding the limit object of a sequence of energies $e^{\veps}$ on
rescaled lattice models is one that has been widely used and
justified in, for example, \cite{mielke05,francfort04,ls05}. In
\cite{mielke05}, the Wigner transform of the normal modes is
employed in solving the macroscopic transport of energy in the
above harmonic systems for deterministic initial data.  
The same system is considered in \cite{ds05} 
with random initial data and in a larger function space, however, 
excluding the type of concentration effects we study here.

The main use of the Wigner transform is that, unlike the energy
density $e^\vep$ itself, it contains enough information so that it
satisfies a closed equation in the limit $\vep \to 0$.
Indeed, it was shown in \cite{mielke05} that,
as long as there is no concentration on the
singular set of the dispersion relation faster than $\vep^{1/2}$, the
Wigner transform of the time-evolved state vector $\psi(t/\vep)$
converges to a limit measure $\tilde\mu_t$ on $\R\times \K$. Here $\K$ is a
suitable compactification of $\T\setminus \Sbb$, $\Sbb$ $=$ the singular
set, which allows a continuous
extension of the group velocity $\nabla \omega$.  The measure $\tilde\mu_t$ is
then proven to satisfy the transport equation
\begin{align}\label{eq:tildemut}
\partial_t \tilde\mu_t(x,k) +  \tfrac{1}{2\pi}
\nabla \omega(k) \cdot \nabla_{\!x} \tilde\mu_t(x,k) =0 .
\end{align}
It was also shown in \cite{mielke05}, that if the weak limit of the rescaled
initial data exists, then the limit satisfies the continuum wave equation.
However, removing the assumption about the rate of concentration, as well
as combining the result with a non-zero weak limit for the initial data
remain open questions.

In this paper we will show how to overcome the above difficulties in a
physically relevant class of models
with a singular dispersion relation.
In Section~\ref{sec:micromodel} we will first present the
microscopic dynamical model in detail.  In
Section~\ref{sec:energy} we will define the Wigner transform, and
discuss its relation to the energy of the microscopic lattice
model.  The main results will be presented in
Section~\ref{sec:mainres}.

\subsection{Relation with solid state physics}
\label{sec:physics}

A crystal in solid state physics is a state of matter in which
the atoms retain a nearly perfect periodic structure over macroscopic times.
The Hamiltonian model used for the
time-evolution in such a crystal is, to the first order accuracy, harmonic.
If we assume that each periodic cell of the idealized perfectly periodic crystal
structure contains $n$ atoms, then we can form a 
vector $q(\gamma)\in \R^{3 n}$ out of the displacements of the atoms in the
periodic cell labeled by $\gamma\in \Z^3$.  The (classical) Hamiltonian
equations of motion of this harmonic model are then
\begin{align}%\label{eq:}
\dot q_{i}(\gamma,t) = \frac{1}{m_i} p_{i}(\gamma,t), \qquad
%\nonumber \\
 \dot p_{i}(\gamma,t) = -\sum_{\gamma',i'}
 \Fmatrix(\gamma-\gamma')_{i,i'} q_{i'\!}(\gamma',t) 
\end{align}
where $\gamma\in \Z^3$, $i=1,\ldots,3 n$, and $m_i$ denotes the mass of
the atom whose displacement $q_i$ measures.

By the change of variables to   
$\tilde{q}_i(\gamma) = m_i^{\frac{1}{2}} q_i(\gamma)$,
$\tilde{p}_i(\gamma) = m_i^{-\frac{1}{2}} p_i(\gamma)$,
these equations can be transformed into a standard form whose 
force matrix is given by
$\tilde{\Fmatrix}(\gamma)_{i,i'}=
m_i^{-1/2} \Fmatrix(\gamma)_{i,i'} m_{i'}^{-1/2}$.
The standard form equations can then be solved by 
Fourier transform, and a diagonalization of the remaining
multiplicative evolution equations decomposes the $3 n$ vector degrees
of freedom into independent {\em normal modes\/}, called 
{\em phonons\/} in solid state physics. Each normal mode is a complex 
scalar field on the crystal lattice, and its time-evolution is
unitary and uniquely determined by the corresponding dispersion relation 
$\omega_i(k)$ on $\T$.  More details about the related mathematical issues
can be found in \cite{mielke05,ds05}. 

From the physical perspective it is highly relevant to understand 
energy transport in the case where $\omega$ is acoustic in
the sense that it contains a $|k|$ singularity. In this case
$\omega$ is degenerate for large wave-numbers in the sense that
the velocity at which long waves propagate converges to the speed
of sound. In other words, the energy density can travel
ballistically over large distance without experiencing dilution
effects due to dispersion.

Acoustic dispersion relations arise in general from atomistic
Hamiltonians with short range harmonic interactions.
For instance,
for the type of interactions considered in \cite{mielke05} all
dispersion relations are Lipschitz continuous and piecewise
analytic. In solid state physics, the modes are accordingly
divided into {\em optical\/} and {\em acoustic\/} depending on
this regularity: if the dispersion relation is regular at $k=0$,
then the mode is called optical, if it behaves as $|k|$ at $k=0$,
then the mode is called acoustic. The latter name arises as these
modes are believed to be responsible for the propagation of sound
waves in the crystal.

The simplicity of the Fourier-picture is the reason that the
dispersion relation is the starting point of most of the more
physically oriented publications. Obviously, within the context of
harmonic defect-free crystals both approaches are completely
equivalent. In the beginning of Section~\ref{sec:micromodel} we
will provide the mathematical details which establish this
equivalence on an elementary level.

Finally, let us remark that the discrete {\em linear} wave equation
alone does not suffice to determine the physically relevant
properties of the crystal, such as its thermal conductivity.
However, it forms the basis for perturbative treatments using which these
questions can be addressed. We refer to
%\cite{ashcroft:solid,Ziman67,spohn05,spohn05err,spohn06}
\cite{ashcroft:solid,Ziman67,spohn05} 
for further details on the
physical aspects of the topic and to \cite{bonetto00} for a review
about related open problems.

\section*{Acknowledgments}

We would like to thank Alexander Mielke and Herbert Spohn for several
instructive discussions.
JL was supported by the
Deutsche Forschungsgemeinschaft (DFG) projects SP~181/19-1
and SP~181/19-2.

\section{The microscopic model} \label{sec:micromodel}

There are two mathematically equivalent descriptions of harmonic
crystals. On the one hand, one can work with the Hamiltonian
equations of motion and analyze the properties of the solutions.
At least formally anharmonic crystals can be discussed in the same
way. On the other hand, one can utilize the periodicity and
linearity to condense the Hamiltonian into the dispersion
relation. While this approach is very elegant, it cannot be used
to directly analyze nonlinear models.

We will show in this chapter how for harmonic lattices the first
approach reduces to the second one. Then we demonstrate how the
tool of Wigner functions can be brought to bear on the reduced
description.

\begin{figure}
%\hspace*{-1.5cm}\includegraphics[height=6cm,width=20cm]{snapshots_section.eps}
\hspace{-1.8cm}\includegraphics[width=18cm]{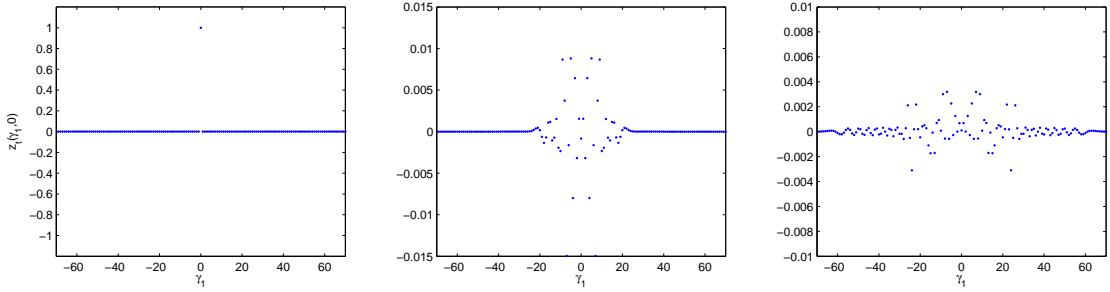}
\caption{Values of $u(\gamma,t)$ along the axis
$\gamma_2=\gamma_3=0$
  for $t=0$, $t=0.1/\vep$, $t=0.9/\vep$ with
  $\vep=\frac{1}{7}*10^{-1}$. The evolution is given by
  (\ref{eq:defWaveeq}) with the nearest neighbor elastic couplings
  (\ref{eq:alphann}), and the initial conditions are $u_{t=0}(0) = 1$,
  $u_{t=0}(\gamma) = 0$ for all $\gamma \not=0$ and $\dot
  u_{t=0}\equiv 0$. \label{fig2}}
\end{figure}

We assume that the scalar excitation $u_t(\gamma)$, $\gamma \in
\Z^3$ satisfies the discrete wave equation 
\begin{equation}\label{eq:defWaveeq} 
\frac{\partial^2}{\partial t^2} u_t(\gamma) =
-\sum_{\gamma'\in \Z^3}\alpha(\gamma-\gamma')u(\gamma')
\end{equation} 
with initial data $(u|_{t=0},\partial_t u|_{t=0}) \in
X = \ell_2 \times \ell_2$. The numbers $\alpha(\gamma-\gamma')$
are the elastic coupling constants between the sites $\gamma$ and
$\gamma'$.  We assume that $\alpha$ is real and symmetric
($\alpha(-\gamma)=\alpha(\gamma)$). Clearly system
(\ref{eq:defWaveeq}) can be written in a Hamiltonian form and the energy
\begin{equation} \label{eq:defHtot1} E(u) = H(u,\dot u) =
\frac{1}{2} \left( \sum_{\gamma \in \Z^3} |\dot u(\gamma)|^2 +
\sum_{\gamma \in \Z^3} \left[ \sum_{\gamma' \in \Z^3}
u(\gamma)\alpha(\gamma-\gamma')u(\gamma')\right]\right)
\end{equation}
is constant along solutions. Depending on the initial conditions
the solutions of system (\ref{eq:defWaveeq}) may develop large
scale oscillations which carry a finite amount of energy, cf.\
Fig.~\ref{fig2} where  snapshots of $u$ at several times are
plotted.

Since system (\ref{eq:defWaveeq}) is linear and invariant under
discrete translations, we can write the solutions in a closed form
using the Fourier transform.
\begin{defn}\label{th:deffourier}
We define the Fourier transform $\ell_2(\Z^3)\to L^2(\Tt)$ by
extending
\begin{align}\label{eq:defgtok}
(\fourier_{\gamma\to k} \psi)(k) =
\hat{\psi}(k) = \sum_{\gamma\in \Z^3} \E^{-\tpi k \cdot \gamma} \psi(\gamma)
\end{align}
from $\psi$ with finite support to all of $\ell_2(\Z^3)$.  Here
$\Tt=\R^3/\Z^3$ denotes the unit 3-torus.   The inverse
transform is pointwise convergently defined by the integral
\begin{align}%\label{eq:}
(\fourier_{k\to \gamma} \hat \psi)(\gamma) =
\int_{\Tt}\D k\, \E^{\tpi k \cdot \gamma} \hat\psi(k)  = \psi(\gamma).
\end{align}
where the measure $\dd k$ is induced by the
Lebesgue measure on $[-\frac{1}{2},\frac{1}{2}]^3$.   In particular,
$\int_{\T} \dd k=1$.
\end{defn}
If one applies the Fourier transform to equation
(\ref{eq:defWaveeq}) one obtains the simpler system
\begin{align}\label{eqn3}
 \frac{\partial}{\partial t}
\begin{pmatrix} \hat u(k,t) \\ \hat v(k,t)
\end{pmatrix} = \begin{pmatrix}
  0 & 1\\ -\omega^2(k) & 0\end{pmatrix}
\begin{pmatrix} \hat u(k,t) \\ \hat v(k,t)
\end{pmatrix}.
\end{align}
The function $\omega:\T \to \R$ is the dispersion relation and is
related to the atomistic Hamiltonian via the following formula
\begin{align}\label{eq:dispersion} 
\omega(k) = \sqrt{\hat \alpha(k)} = \sqrt{\sum_{\gamma \in \Z^3}
\alpha(\gamma)\cos(2 \pi \gamma\cdot k)}.
\end{align}
where we have employed the assumption $\alpha(\gamma) = \alpha(-\gamma)$.
Since $\alpha$ is real and satisfies the above symmetry property,
we find that $\omega$ is also
real and symmetric, i.e. $\omega(k) = \omega(-k)$.

Now diagonalizing the matrix on the right hand side of (\ref{eqn3})
motivates combining the real scalar fields $u,v$ into the two complex fields
$\psi_\pm=\psi_\pm[u,v]\in \ell_2(\Z^3,\C)$ defined by the formula
\begin{align}\label{eq:defpsisigma}
\hat{\psi}_{\sigma}(k)=\frac{1}{\sqrt{2}}\left(  \omega(k)
\hat{u}(k)
 + \I \sigma  \hat v(k) \right)
\end{align}
where $\sigma = \pm 1$.  For all $(u,v)\in X$, we clearly have
$\hat{\psi}_{\sigma}\in L^2(\T)$, and thus $\psi_{\sigma}\in
\ell_2(\Z^3)$. In addition, since we assumed
$\omega(-k)=\omega(k)$, we also have
$\psi_{-}(\gamma)=\overline{\psi_{+}(\gamma)}$ for all $\gamma$.
The transformation can always be inverted by applying
\begin{align}%\label{eq:}
  \hat v = -\frac{\I }{\sqrt{2}} (\hat\psi_+ - \hat\psi_-),
\quad   \hat u = \frac{1}{ \omega\sqrt{2}} (\hat\psi_+ +
\hat\psi_-).
\end{align}
These fields are normal modes of the harmonic system, since
(\ref{eqn3}) implies that $\psi_\pm(\gamma,t)=
\psi_\pm[u(t),v(t)](\gamma)$ satisfy the evolution equations
\begin{align}%\label{eq:}
 \frac{\partial}{\partial t}
\begin{pmatrix} \hat \psi_+(k,t) \\ \hat \psi_-(k,t)
\end{pmatrix}
= -\I \begin{pmatrix}
  \omega(k) & 0 \\ 0 &  -\omega(k) \end{pmatrix}
\begin{pmatrix} \hat \psi_+(k,t) \\ \hat \psi_-(k,t)
\end{pmatrix}  ,
\end{align}
which are readily solved to yield for all $t\in \R$, $k\in \R$
\begin{align}\label{eq:modesol}
 \hat \psi_{\pm}(k,t) = \E^{\mp \I \omega(k) t} \hat \psi_{\pm}(k,0) \,.
\end{align}
These are exactly the two evolution equations corresponding to a
``phonon'' mode with a dispersion relation $\omega$.

After these reduction steps it is obvious that the dispersion
relation $\omega$ fully determines the properties of the solutions. 
We will assume throughout this paper that $\omega$ is of
acoustic type in the following precise sense:
\begin{defn}\label{th:defomega}
We call $\omega \in C(\Tt, [0,\infty))$ an
{\em acoustic dispersion relation\/} if $\lambda=\omega^2$ satisfies:
\begin{enumerate}
\item $\lambda\in C^{(3)}(\Tt,[0,\infty) )$.
\item $\lambda(0)=0$, and the Hessian of $\lambda$ is invertible at $0$.
\end{enumerate}
A dispersion relation is called {\em regular acoustic,\/} if it is
acoustic and $\lambda(k)>0$ for $k\ne 0$.  The $3\times 3$-matrix
$A_0$ is the Hessian of $\frac{1}{2}\lambda$ at $k=0$ and
$\omega_0(q) =\sqrt{q \cdot A_0 q}$.
\end{defn}
These assumptions are fairly general: as discussed in the
introduction, all stable harmonic interactions have non-negative
eigenvalue functions $\lambda$, and for interactions of the type
discussed in \cite{mielke05} $\omega=\sqrt{\lambda}$ is Lipschitz
continuous.

A prototype for the kind of dispersion relations we will consider
here is the dispersion relation of the nearest neighbor square
lattice:
\begin{align}\label{eq:nndisp}
 \omega_{\text{nn}}(k) = \Bigl[\sum_{\nu=1}^3
    2(1-\cos(2 \pi k^{\nu}))\Bigr]^{\frac{1}{2}}.
\end{align}
This is clearly a regular acoustic dispersion relation, in the
sense of Definition \ref{th:defomega}, and for it $A_0$ is
proportional to a unit matrix and $\omega_0(q)=2 \pi|q|$. The
corresponding elastic couplings are given by
$\alpha_{\text{nn}}(\gamma'-\gamma) = - \Delta_{\gamma'\gamma}$,
where $\Delta$ is the discrete Laplacian of the square lattice.
Explicitly,
\begin{align}\label{eq:alphann}
  \alpha_{\text{nn}}(\gamma) = \begin{cases}
    6 , & \text{if } \gamma=0,\\
    -1 , & \text{if } |\gamma|=1,\\
    0 , & \text{otherwise.}
\end{cases}
\end{align}
To allow the creation of macroscopic oscillations we work with
sequences of initial conditions that depend on the scaling-parameter
$\veps>0$ and consider the asymptotic behavior of the solutions as
$\veps$ tends to 0. 

% A typical initial condition might be
% $\psi_+^\veps(\gamma,0) = \veps^{-3/2} \int_{[-\frac{1}{2},\frac{1}{2}]^3}
% \sin(2 \pi k\cdot \gamma)
% f(\veps\gamma + k)\, \dd k$ where $f \in L^2(\R^3)$. 

% From now on we will neglect the sign factor
% $\sigma$ and consider only the case $\sigma=+1$.  The results can then be
% extended to the mode $\sigma=-1$ by replacing $\omega$ by $-\omega$
% wherever it appears.

\subsection{Energy density and the lattice Wigner transform}
\label{sec:energy} 

From now we will focus on analyzing asymptotic
behavior of the fields $\psi_\pm$ as $\veps$ tends to 0. As is
carefully discussed in \cite{mielke05}, generalizing the
definitions of the energy density and of the Wigner transform to
the discrete setting is not completely obvious.  In an attempt to
minimize unnecessary repetition of certain basic results related
to Wigner transforms, we will resort here to the definitions used
in \cite{ls05} which will allow us to rely on the properties
proven in Appendix B of that reference. However, we wish to keep
in mind that this choice might not be optimal for all purposes,
and we refer the interested reader to the discussion and to the
references in \cite{mielke05,SP04} for further possibilities.

We employ here the definition that for any state $(u,v)$, its energy
density, $e^\vep=e^\vep[u,v]$, scaled to a lattice spacing $\vep>0$,
is the tempered distribution defined via the complex fields
$\psi_\sigma=\psi_\sigma[u,v]$ in (\ref{eq:defpsisigma}):
\begin{align}%\label{eq:}
e^\vep(x) = \sum_{\gamma\in \Z^3} \delta(x-\vep \gamma)
\frac{1}{2} \sum_{\sigma=\pm 1} |\psi_\sigma(\gamma)|^2
\end{align}
where $\delta$ denotes the Dirac delta-distribution.
This is a manifestly positive distribution, and identifiable with
a measure whose total mass equals the total energy:
\begin{align}%\label{eq:}
&\int \dd x\, e^\vep(x) = \sum_{\gamma\in \Z^3}\frac{1}{2}
\sum_{\sigma=\pm 1} |\psi_\sigma(\gamma)|^2 = \frac{1}{2}
\sum_{\sigma=\pm 1} \norm{\psi_\sigma}^2 \nonumber \\ & \qquad =
\frac{1}{4} \sum_{\sigma=\pm 1} \int\dd k \left|  \omega(k)
\hat{u}(k) + \I \sigma  \hat v(k) \right|^2 = H(u,\dot u)<\infty.
\end{align}
This justifies calling $e^\vep$ an energy density: it defines a
distribution of the positive total energy between the lattice
sites. The symmetry of $\omega$ implies that
$|\psi_{-}(\gamma)|=|\psi_{+}(\gamma)|$ and thus we can also identify
the energy density directly with the norm-density of $\psi_+$:
\begin{align}%\label{eq:}
e^\vep(x) = \sum_{\gamma\in \Z^3} \delta(x-\vep \gamma)
|\psi_+(\gamma)|^2.
\end{align}
Then also for all $\vep>0$,
\begin{align}%\label{eq:}
H(u,v)= \int \dd x\, e^\vep(x) =
\|\psi_+\|^2_{\ell_2(\Z^3)}=\|\hat \psi_+\|_{L^2(\Tt)}^2
= \|\hat \psi_-\|_{L^2(\Tt)}^2
\end{align}
which is conserved when $(u,v)=(u(t),v(t))$. 

\begin{figure}
\begin{center}
%\hspace*{-1.5cm}
\includegraphics[height=12cm]{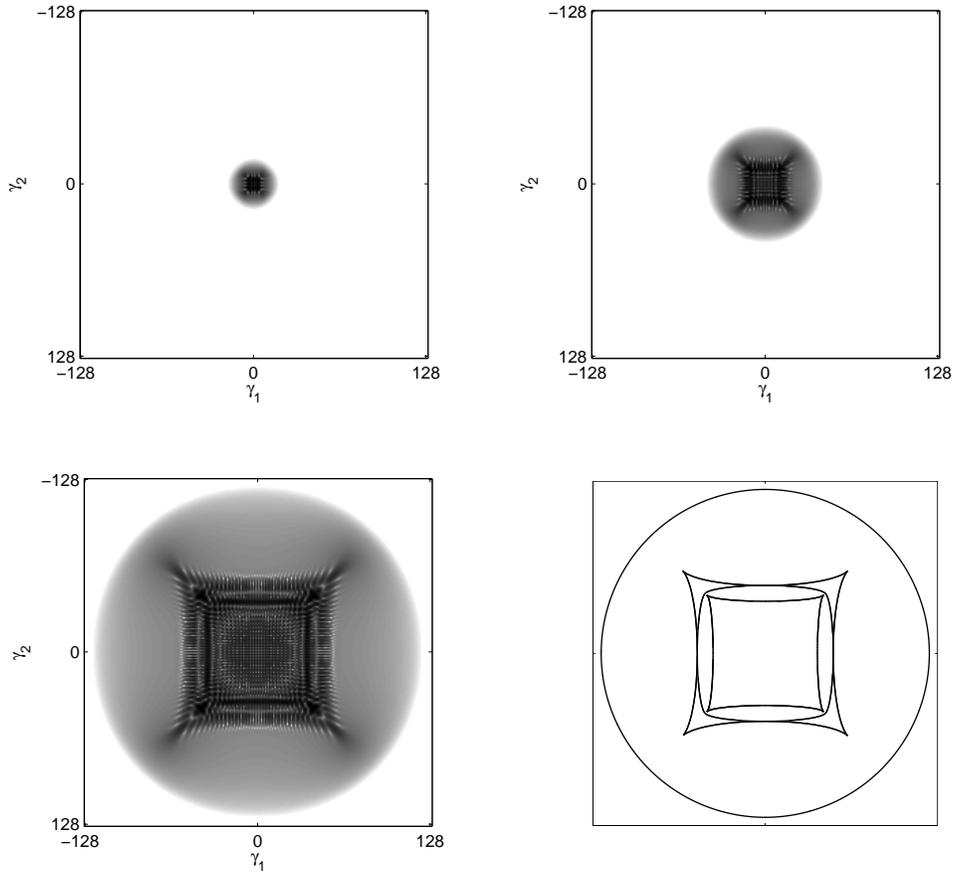}
\end{center}
\caption{First three panels: Snapshots of the energy density
$|\psi_+(\gamma,t)|^2$ in the plane $\gamma_3=0$ at
$t=0.1/\vep$, $t=0.3/\vep$ and
$t=0.95/\vep$, $\vep=\frac{1}{128}$, with initial data and elastic
constants as in Fig.~\ref{fig2}.  The plot is of the logarithm of the
density, with all values less than a fixed cut-off shown white.  
Last panel: Plot of the restriction  to the
plane $x_3=0$ of the 
singular set of the solution to the transport equation (\ref{tran1})
with the corresponding macroscopic initial data.
As the solution is scale-invariant, no explicit length scale has been
denoted.  
\label{fig1}}
\end{figure}

We have given an
example of the time-evolution of the so defined energy density in
Fig.~\ref{fig1}.  The last panel in the figure 
contains the most obvious features
which are implied by the corresponding macroscopic evolution equation:
the points of discontinuity of the solution.  The macroscopic initial data
is given by 
$\mu_0(\D x,\D k) = \delta(x) \frac{1}{2}|\omega(k)|^2 \D x \D k$, which
has no concentration at $k=0$.  Thus only (\ref{tran1}) is relevant.
It is readily solved to yield as the energy density
\begin{align}%\label{eq:}
e(x,t) = \int_{\T} \D k\,\delta(x-t\tfrac{1}{2\pi}\nabla \omega(k))
 = t^{-3} \int_{\T} \D k\, \delta(\tfrac{x}{t}-\tfrac{1}{2\pi}\nabla \omega(k))
\end{align}
Evaluating such integrals has been considered, for instance, in Sec.~6.4 of
\cite{mielke05}.  $|\nabla \omega(k)|$ has its maximum near the point of
discontinuity of the gradient, at $k=0$.  This defines the outer circle
outside which the solution must be zero.  Inside the circle, the solution
has a finite density, apart from points which correspond to values of $k$
for which the Hessian of $\omega$ is not invertible.  We have
computed the positions of such points using Mathematica, and plotted the
result in the last panel in Fig.~\ref{fig1}.  For a reader interested in
the details of the computation, we point out that considering the case 
$x_3=0$ simplifies the problem, as it implies that either $k_3=0$ or
$k_3=\frac{1}{2}$.

We are interested in the limiting behavior of
$e^\vep[u(t/\vep),\dot u(t/\vep)]$, as $\vep$ tends to 0. Since
the velocity of the waves with wave vector $k$ depends on $k$ it
is necessary to work with an object that encodes the density of
waves with wave vector $k\in \Tt$ at $x \in \R^3$. This job is
conveniently done by the Wigner-transform. In order to avoid
certain technical difficulties we are going to define our
Wigner-transform only in the sense of distributions, i.e., via a
duality principle.

First we introduce the space of Schwartz functions.
\begin{defn}\label{th:defSnorm}
Let $\schwartz_d=\schwartz(\R^d)$ denote the Schwartz space,
and $\Sdnorm{\cdot}{d}{N}$ the corresponding $N$:th Schwartz norm.
Explicitly, with $\alpha$ denoting an arbitrary multi-index
and with $\sabs{x}=\sqrt{1+x^2}$, then
\begin{align}%\label{eq:}
 \Sdnorm{f}{d}{N} = \sup_{x\in \R^d}\max_{|\alpha|\le N}
 |\sabs{x}^N \partial^\alpha f(x)| .
\end{align}
We also employ the shorthand notation $\schwartz=\schwartz_3$.
\end{defn}
To extract the relevant weak limits from the sequence $\psi^\vep$
as $\vep$ tends to 0 we have to specify a space of suitable
test functions. Since we want to track the evolution of three
different kinds of lattice vibrations (short-, medium- and
long-wavelength) we require a somewhat involved and non-standard notion
of multiscale test functions.
\begin{defn}\label{th:deftestf}
We call a test function $a\in C^{(\infty)}(\R^3\times \Tt \times \R^3)$
{\em admissible\/}, if it satisfies the following properties:
\begin{enumerate}
\item $\sup_{k,q,|\alpha|\le N}
  \Snorm{\partial^\alpha a(\cdot,k,q)}{N}<\infty$, for all $N\geq 0$,
\item $q\mapsto a(x,k,q)$ is constant for all $\mod{k}_\infty \ge
  \frac{1}{4} $ and $x\in \R^3$.
\item\label{it:defbfunc}
  There is a function $b\in C^{(\infty)}(\R^3\times \Tt \times S^{2})$
  such that for any $N\ge 0$
  \begin{align}%\label{eq:}
    \sup_{|q|\ge R,k\in\Tt}
    \norm{a(\cdot,k,q)-b(\cdot,k,\tfrac{q}{|q|})}_{\schwartz,N}\to
    0,\quad \text{when }R\to\infty.
  \end{align}
\end{enumerate}
\end{defn}
The first condition can be summarized as follows:
we assume the test-functions to be
Schwartz in $x$ and smooth with bounded derivatives in $k$ and $q$.
The above requirements are not minimal.
The second condition is only needed in order to guarantee that
$k\mapsto a(x,k,k/\vep)$ would always be smooth on $\T$.
Also, taking arbitrarily large $N$ in the last step is not necessary, most
likely  $N=d+3=6$ would suffice.

Having the notion of admissible test functions at our disposal we can define
the central object of this paper: the Wigner transform.
\begin{defn} \label{def:Wigners}
Let $\psi \in \ell_2(\Z^3)$.
We define the lattice Wigner transform $\Wep[\psi]$ at scale
$\veps>0$ by
\begin{align}\label{eq:defLW}
\mean{a,\Wep[\psi]}% \nonumber \\ & \quad
=  \int_{\R^3}\dd p\,
\int_{\Tt}\dd k\, \overline{\hat a(p,k,\tfrac{k}{\vep})} \, \overline{\hat
\psi(k-\veps\tfrac{p}{2})} \hat \psi(k+\veps\tfrac{p}{2}),
\end{align}
where $a$ is an admissible test function and 
$\hat a = \fourier_{x\to p} a$, i.e., 
\begin{align}%\label{eq:}
\hat{a}(p,k,q) = \int_{\R^3} \D x\, \E^{-\tpi p \cdot x} a(x,k,q).
\end{align}
The $L^2$-Wigner transform $\Wcont^{(\veps)}[\phi]$ of a function $\phi \in
L^2(\R^3)$ at the scale $\veps>0$ is given by the distribution
\begin{align}\label{eq:defEWcont}
b\mapsto \mean{b,\Wcont^{(\veps)}[\phi]} =
  \int_{\R^3 \times \R^3}\! \D p\,\dd q\, \overline{\hat b(p,q)}\,
\overline{\hat \phi(q-\veps\tfrac{p}{2})} \hat
\phi(q+\veps\tfrac{p}{2})
\end{align}
for all $b\in \schwartz(\R^3,C^{(\infty)}(\R^3))$, and with $\hat b =
\fourier_{x\to p} b$.
\end{defn}
The test-function space $\schwartz(\R^3,C^{(\infty)}(\R^3))$ used above
to define $\Wcont$ is obtained
via the family of seminorms
$p_N(b)= \sup_{|\alpha|,|x|,|q|\le N}|\sabs{x}^N D^\alpha b(x,q)|$
with $b\in C^{(\infty)}(\R^3\times \R^3)$.
This is a \frechet -space, and $\Wcont[\phi]$ is a continuous functional on
it for any $\phi\in L^2(\R^3)$, as the following estimate reveals:
\begin{align}\label{eq:Wcontest}
  | \mean{b,\Wcont^{(\veps)}[\phi]} | \le
  \sup_{p,q}{|\sabs{p}^{4} \hat{b}(p,q)|}\,
 \norm{\phi}^2 \int_{\R^3}\!\D p\, \sabs{p}^{-4}.
\end{align}
Although $\schwartz_6$ is not dense in this test-function space, it is
nevertheless enough to know how $\Wcont$ acts on it.  More precisely, if
$W_i=\Wcont^{(\veps)}[\phi_i]$, $i=1,2$, and
$\mean{b,W_1}=\mean{b,W_2}$ for all $b\in \schwartz_6$, then
$W_1=W_2$.  This follows straightforwardly from
an estimate similar to (\ref{eq:Wcontest}) using smooth
cutoff functions to cut out the infinity of the $q$-variable.
In addition, we will also need the property that, if $b(x,q)=f(x)$,
with $f\in \schwartz_3$, then
\begin{align}%\label{eq:defEWcont}
\mean{b,\Wcont^{(\veps)}[\phi]} =
  \int_{\R^3}\! \D x \overline{f(x)} |\phi(x)|^2.
\end{align}
That is, at least formally,
$\int_{\R^3}\dd q\, \Wcont^{(\veps)}[\phi](x,q) = |\phi(x)|^2$.

In \cite{ls05} the Wigner transform of a lattice state
was defined as a distribution
$W_{\text{latt}}^\vep \in \schwartz'(\R^3\times \T)$.  The above
definition is simply a refinement of this definition: formally
for any $\psi$
\begin{align}%\label{eq:}
  \Wep(x,k,q) = \delta\bigl(q-\tfrac{k}{\vep}\bigr) W_{\text{latt}}^\vep(x,k).
\end{align}
This follows immediately from Eq.~(B.6) of \cite{ls05}, after one realizes
that if $a$ is an admissible test function, then
$(x,k)\mapsto a(x,k,k/\vep)$ belongs to $\schwartz(\R^3\times \T)$
for any $\vep>0$.
This identification immediately allows us to use the results in
\cite{ls05} and to prove that many of the basic properties of the usual
Wigner transform carry over to
the multi-scale Wigner transform.  Particularly important for us is the
following relation:
\begin{prop}
For any $f\in \schwartz(\R^3)$, the test function
$a_f(x,k,q) = f(x)$
is admissible, and for all $\vep>0$ and for all $(u,v)\in X$,
\begin{align}\label{eq:enrel}
  \mean{f,e^\vep} = \mean{a_f,\Wep[\psi]}
\end{align}
where $e^\vep=e^\vep[u,v]$ and $\psi=\psi_+[u,v]$.
\end{prop}
\begin{proof}
By inspection, we find that $a$ is a well-defined test-function, and
$\psi\in \ell_2$.
Then, by the above mentioned relation,
$\mean{a_f,\Wep[\psi]} = \mean{J_f,W_{\text{latt}}^\vep[\psi]}$
with $J_f(x,k)=f(x)$.
On the other hand, it follows directly from the definition of
$W_{\text{latt}}^\vep$ (equation (B.2) in \cite{ls05}) that
\begin{align}%\label{eq:}
& \mean{J_f,W_{\text{latt}}^\vep[\psi]} =
\sum_{\gamma,\gamma'\in \Z^3} \psi(\gamma) \overline{\psi(\gamma')}
 \int_{\T}\dd k\, \E^{2 \pi \I k \cdot (\gamma'-\gamma) }
   \overline{f(\vep (\gamma'+\gamma)/2)}
\nonumber \\ &  \quad
=  \sum_{\gamma\in \Z^3} \overline{f(\veps \gamma)} |\psi(\gamma)|^2 = \mean{f,e^\vep}.
\end{align}
This proves (\ref{eq:enrel}).
\end{proof}

The above Proposition can be formally summarized by the formula
\begin{align} \label{kmargin}
e^\vep(x) = \int_{\Tt}\dd k\int_{\R^3}\dd q\, \Wep[\psi](x,k,q)
\end{align}
which implies also
(in the sense of choosing any suitable test-function sequence
approaching pointwise $1$)
\begin{equation} \label{normeq}
\int_{\R^3 \times \Tt\times\R^3}\dd x\,\dd k\,\dd q\, \Wep[\psi](x,k,q) =
\|\psi\|^2_{\ell_2(\Z^3)}.
\end{equation}
As noted earlier, analogous results hold for the $L^2$-Wigner transform.
Let us also remark here that, if the sequence
$W_{\text{latt}}^\vep[\psi^\vep](\cdot,\cdot)$  converges, then the limit
is given by a non-negative Radon-measure \cite{ls05}.
The results of the next section will show that
% not only $\Wep[\psi^\vep](\cdot,\cdot)$ converges to a non-negative
% Radon-measure, but
also $W_{\text{latt}}^\vep[\psi^\vep](\cdot,\cdot/\vep)$ has such a
limit property.

\section{Main results}
\label{sec:mainres}

\label{sec:semi} The macroscopic evolution is obtained by sending
$\veps$ to 0. Our objective is to characterize the asymptotic
behavior of the Wigner function $\Wep[\psi^\vep]$. The limit
strongly depends on the dispersion relation $\omega$.
We will consider here regular acoustic dispersion relations, keeping in
mind that in
the case of a scalar field and nearest neighbor interactions in
$\Z^3$ the dispersion relation $\omega$ is given by (\ref{eq:nndisp}) which
is regular acoustic with $\omega_0(q)=2\pi |q|$.
The main achievement of this paper is that complicated assumptions
concerning the concentrations of the Wigner transform $\Wep$ in
wave-number space as $\veps$ tends to 0 are no longer needed. The
only remaining requirements are boundedness and tightness of the
sequence of initial excitations.
\begin{assump}\label{th:mainassump}
We consider a sequence of values $\veps>0$ such that $\veps\to 0$.
For each $\veps$ in the sequence we assume that there is given an initial
data vector $\psi^\veps_0\in \ell_2(\Z^3)$ such that
\begin{enumerate}
\item $\sup_\veps \norm{\psi^\veps_0} <\infty$.
\item The sequence $\psi^\veps_0$ is tight on the scale $\veps^{-1}$:
  \begin{align}%\label{eq:}
\lim_{R\to\infty}
  \limsup_{\veps\to 0} \sum_{|\gamma|> R/\veps}
  |\psi^\veps_{0}(\gamma)|^2 = 0.
  \end{align}
\end{enumerate}
\end{assump}
After these preparations we are in a position to state our result.
The main point is that if $\omega$ is a regular acoustic
dispersion relation the asymptotic behavior of the energy density
is characterized by precisely three different objects: the weak
limit (macroscopic waves), the H-measure (short macroscopic waves)
and the Wigner measure (microscopic waves) and no assumption
concerning energy concentrations except those stated in
Assumption~\ref{th:mainassump} are required.
\begin{thm} \label{th:mainres}
Let $\psi_0^\vep \in \ell_2(\Z^3)$ be a sequence which satisfies
Assumption \ref{th:mainassump}.
Let $\omega$ be a regular acoustic
dispersion relation and define $\psi^\veps_t\in \ell_2(\Z^3)$ for
all $t\in \R$ by the formula
\begin{align}%\label{eq:}
  \hat{\psi}_t^\veps(k) = \E^{-\I t \omega(k)} \hat{\psi}^\veps_0(k).
\end{align}
Let also $\T_*=\T\setminus\{0\}$.
Then there are positive, bounded Radon measures $\mu_0, \muH_0$ on
$\R^3\times \Tt_*$ and $\R^3\times S^2$, respectively, a function
$\phi_0 \in L^2(\R^3)$ and a subsequence (not relabeled) such
that for all admissible test functions $a$ and $t\in \R$,
\begin{align} \label{eq:entr}
\lim_{\veps\to 0} \mean{a, \Wep[\psi^\veps_{t/\vep}]}=&
\int_{\R^3 \times \T_*} \dd \mu_t(x, k)\,
\overline{b(x,k,\tfrac{k}{|k|})}\,
\\\nonumber& + \int_{\R^3 \times S^2} \dd \muH_t(x,q)\,
\overline{b(x,0,q)}\, + \mean{a_0,\Wcont^{(1)}[\phi_t]}\,,
\end{align}
where $b(x,k,q) = \lim_{R\to\infty} a(x,k,R q)$ for $|q|=1$,
$a_0(x,q)=a(x,0,q)$, and $\phi_t$, $\mu_t$ and $\muH_t$ are given
by
% $\hat \phi(q;t) =\E^{\I t \omega_0(q)} \hat{\phi}_0(q)$.
\begin{eqnarray}%\label{eq:}
\hat \phi_t(q) &:=&\E^{-\I t \omega_0(q)} \hat{\phi}_0(q),\\[4mm]
\int_{\R^3\times \T_*} \, a(x,k)\,\D\mu_t(x,k)
&:= &\int_{\R^3\times \T_*} \,
a(x+t\tfrac{1}{2\pi}\nabla \omega(k),k)\,\D\mu_0(x ,k),
\label{eq:defmut}\\[1mm]
\int_{\R^3 \times S^2} \, b(x,q)\,\D\muH_t(x ,q)& :=&
\int_{\R^3 \times S^2} \, b(x+t\tfrac{1}{2\pi}\nabla
\omega_0(q),q)\,\D\muH_0(x,q)\,. \label{eq:defmuht}
\end{eqnarray}

Moreover, for all $t$ the energy equality
\begin{equation}
\label{eq:eneq} \lim_{\veps\to 0} \|\psi^\veps_0\|^2 =
 \mu_t(\R^3\times \T_*)+ \muH_t(\R^3\times S^2)
+ \norm{\phi_t}_{L^2(\R^3)}^2
\end{equation}
holds.
\end{thm}
\begin{remark}
It is immediate from the definition that $\phi$, $\mu$ and $\muH$ are
weak solutions of the set of decoupled linear transport equations
(\ref{tran1}--\ref{tran3}).
\end{remark}
The proof of Theorem~\ref{th:mainres} also shows that the subsequences
which are extracted in the statement of the theorem can be
characterized by a simple condition.
In particular, the initial state of the wave-equation, $\phi_0$,
is determined as the weak-$L^2(\R^3)$ limit of the sequence of the
functions $(\phi^\veps_0)$ with Fourier-transforms
\begin{align} \label{eq:defu}
\hat{\phi}^\veps_0(q) = \begin{cases} \veps^{\frac{3}{2}}
  \hat{\psi}^\veps_0(\veps q)& \text{if } \norm{q}_\infty\le
\frac{1}{2\veps},\\
0 & \text{otherwise.}\end{cases}
\end{align}
The exact characterization is contained in the following Corollary whose
proof will be given in Section \ref{sec:corrproof}.
\begin{cor}\label{th:maincorr}
Let $(\psi^\vep_0)$ be a sequence which satisfies
Assumption~\ref{th:mainassump}. Suppose that $\phi^\veps_0$
converges weakly to $\phi_0$, and that
$\lim_{\veps\to 0} \mean{a, \Wep[\psi^\veps_0]}$ exists for
every admissible testfunction $a$.
Then there are unique positive, bounded
Radon measures $\mu_0, \muH_0$ on $\R^3\times \Tt_*$ and $\R^3\times
S^2$,
respectively,  such that
for every admissible testfunction $a$
\begin{align} \label{eq:entr2}
\lim_{\veps\to 0} \mean{a, \Wep[\psi^\veps_0]}=&
\int_{\R^3 \times\T_*} \D\mu_0(x ,k)\,
\overline{b(x,k,\tfrac{k}{|k|})}\,
\\\nonumber& + \int_{\R^3 \times S^2} \D\muH_0(x,q)\,
\overline{b(x,0,q)}\, + \mean{a_0,\Wcont^{(1)}[\phi_0]}\,,
\end{align}
where $a_0$ and $b$ are defined as in Theorem~\ref{th:mainres}.
In addition, then (\ref{eq:entr}) holds for all $t\in\R$ along the original
sequence $\vep$ with the initial macroscopic data determined by the triplet
$(\mu_0,\muH_0,\phi_0)$.
\end{cor}

The measure $\muH $ is closely related to H-measures which have
been introduced in the context of oscillatory solutions of partial
differential equations by L.\ Tartar \cite{tartar90} and P.\ G\'erard
\cite{gerard88}. The precise nature of the connection of between
$\muH $ and H-measures is irrelevant for the purpose of this
paper, but for the convenience of the reader we include a brief
discussion.
\begin{defn}[H-measures]
Let $\phi^\veps \in L^2(\R^d)$ be a sequence which converges
weakly to 0 as $\veps \to 0$ and let $\nu \in \mathcal {\mathcal
M}_+(\R^d \times S^{d-1})$ be a nonnegative Radon measure. If
\begin{equation*}
\lim_{\veps \to 0} \int_{\R^d} \overline{{\fourier_{x\to q}}(a_1
\phi^\veps)}
    {{\fourier_{x \to q}}(a_2 \phi^\veps)} \psi(\tfrac{q}{|q|}) \, \dd k =
    \int_{\R^d} \int_{S^{d-1}} \dd \nu(x, q)\, a_1(x) a_2(x) \psi(q)
\end{equation*}
for all $a_1,a_2 \in C^\infty_c(\R^d)$ and $\psi \in C(S^{d-1})$,
then $\nu$ is the H-measure generated by the sequence
$\phi^\veps$.
\end{defn}
The connection between $\muH $ and H-measures is established by
the following
\begin{prop} \label{semcom}
Let $\phi^\veps \in L^2(\R^d)$ be a tight sequence converging
weakly to 0 as $\veps \to \infty$ such that
\begin{equation} \label{nocon}
\lim_{\rho \to 0}\lim_{\veps \to 0} \int_{\veps |k| \geq \rho}\dd
k\, |\hat \phi^\veps(k)|^2 = 0.
\end{equation} If $\phi^\veps$ generates an H-measure $\nu$ as
$\veps \to 0$ and $a$ is an admissible testfunction, then
\begin{equation}
\lim_{\veps \to 0} \int_{\R^3} \dd x\, \int_{\R^3}\dd q\,
\overline{\hat a(p,\veps q, q)}\,
\overline{\hat \phi^\vep(q-\tfrac{p}{2})} \hat \phi^\vep(q+ \tfrac{p}{2})
= \int_{\R^3 \times S^2} \dd \nu(x,q) \, a(x,0,q).
\end{equation}
\end{prop}
\begin{proof} See \cite{Harris07} \end{proof}
Note that we will never assume that $\phi^\veps$ converges weakly
to 0, hence we cannot define the H-measure associated to
$\phi^\veps$. Moreover, unlike the H-measure $\nu$ the measure
$\muH $ does not take the contribution of oscillations with
wavelength $1/\veps$ into account. For these reasons $\muH $
differs from $\nu$ in general.

\section{Proof of Theorem \ref{th:mainres}}
\label{sec:proofofmain}

%\begin{proofof}{Theorem \ref{th:mainres}}
Let $a$ be an admissible testfunction and consider a fixed $t\in
\R$, when we need to inspect the $\vep\to 0$ limit of
\begin{align}\label{eq:defEW}
%& \mean{a,\Wep[\psi^\veps(\tfrac{t}{\veps})]}% \nonumber \\ & \quad
& \mean{a,\Wep[\psi^\veps_{t/\veps}]}% \nonumber \\ & \quad
=  \int_{\R^3}\! \D p\, \int_{\Tt}\! \D k\,
\overline{\hat{a}(p,k,\tfrac{k}{\veps})} \,
\E^{-\I\frac{t}{\veps}
\left(\omega(k +\veps \frac{p}{2})
     -\omega(k -\veps \frac{p}{2})\right)}\,
\overline{\hat{\psi}^\vep_0(k -\veps {\textstyle \frac{p}{2}})}\,
\hat{\psi}^\vep_0(k +\veps {\textstyle \frac{p}{2}})\, .
\end{align}
First we identify the function $\phi_0$ which contains the
contributions of the long-wave excitations. Let
$\hat{\phi}^\veps_0$ be defined by (\ref{eq:defu}).
Since $\limsup_{\veps \to 0} \|\hat \phi^\veps_0\|_{L^2} =
\limsup_{\veps \to 0}\|\psi^\vep_0\|_{\ell_2}$ is bounded by
Assumption~\ref{th:mainassump} there exists a subsequence and a
function $\phi_0 \in L^2(\R^3)$ such that $\phi^\vep_0$ converges
weakly to $\phi_0$ in $L^2(\R^3)$.

Using a localization function $\chi$ which will be specified later
we split the rhs of (\ref{eq:defEW}) into three parts and an
error term so that the contributions short-, medium- and long-wave
excitation can be analyzed separately:
\begin{align*}
\Iouter & = \int_{\R^3}\! \D p\, \int_{\Tt}\! \D k\,
\overline{\hat{a}(p,k,\tfrac{k}{\veps})} \,
\E^{-\I\frac{t}{\veps}\left(\omega(k +\veps \frac{p}{2})
-\omega( k -\veps \frac{p}{2})\right)}\,
\overline{\hat{\psi}^\vep_0(k -\veps {\tfrac{p}{2}})}\,
\hat{\psi}^\vep_0(k +\veps {\tfrac{p}{2}})(1-\chi),\\
\Ismall^{\mathrm{H}} & = \int_{\R^3}\! \D p\, \int_{\Tt}\! \D k\,
\overline{\hat{a}(p,k,\tfrac{k}{\veps})} \,
\E^{-\I\frac{t}{\veps}\left(\omega(
k +\veps \frac{p}{2}) -\omega( k -\veps \frac{p}{2})\right)}\,
\overline{
(\hat{\psi}^\vep_0(k -\veps\tfrac{p}{2})
-\vep^{-\frac{3}{2}}\hat \phi_0(\tfrac{k}{\vep}-\tfrac{p}{2}))}\,\\
& \qquad
\times (\hat{\psi}^\vep_0(k +\veps {\tfrac{p}{2}})- \vep^{-\frac{3}{2}}
\hat\phi_0(\tfrac{k}{\vep} + \tfrac{p}{2}))\chi,\\
\Ismall^{\rm wv}  &= \vep^{-3} \int_{\R^3}\! \D p\, \int_{\Tt}\! \D k\,
\overline{\hat{a}(p,k,\tfrac{k}{\veps})} \,
\E^{-\I\frac{t}{\veps}\left(\omega(
k +\veps \frac{p}{2}) -\omega( k -\veps \frac{p}{2})\right)}\,
\overline{\hat{\phi}_0(\tfrac{k}{\vep} -\tfrac{p}{2})}\,
\hat{\phi}_0(\tfrac{k}{\vep} + \tfrac{p}{2})\chi,\\[2mm]
R &= I - \Iouter-\Ismall^{\mathrm{H}}-\Ismall^{\rm wv}.
\end{align*}
The definition of $R$ implies that
\begin{equation} \label{eq:firstsplit}
\mean{a,\Wep[\psi^\veps_{t/\vep}]}
=\Iouter+\Ismall^{\mathrm{H}}+  \Ismall^{\rm wv}+R.
\end{equation}
To localize the oscillations in Fourier-space we need smooth
cutoff-functions.
\begin{defn}\label{def:cutoff}
Let $\cutoffone\in C^{(\infty)}(\R,[0,1])$ denote a fixed function
which is symmetric, $\cutoffone (-x)=\cutoffone (x)$, strictly
monotonically decreasing on $[1,2]$, and
\begin{align} \label{eq:defvphi}
\cutoffone (x) = \begin{cases} 1& \text{if } |x|\le 1,\\
0 & \text{if } |x|\ge 2. \end{cases}
\end{align}
We define further $\cutoff\in C^{(\infty)}(\R^3,[0,1])$ by
$\cutoff(k)=\cutoffone(|k|)$.
\end{defn}

Let $0<\rho\le \frac{1}{4}$ be arbitrary and set $\chi(k)
=\cutoffrac{k_+}{\rho} \cutoffrac{k_-}{\rho}$ where $k_{\pm} = k
\pm \veps {\textstyle \frac{p}{2}}$. We will continue to use this
shorthand notation, under the tacit assumption that $k_\pm$ is
always really a function of $k$ and $\vep p$. All four terms
depend on $\vep$ and $\rho$, although we do not denote this
explicitly in general.

The first term containing $1-\chi$ in
(\ref{eq:firstsplit}) is zero, if $|k_\pm|\le \rho$, while
the remainder is zero, if $|k_+|$ or $|k_-|\ge 2 \rho$.  Thus the chosen
decomposition splits the integration over $k$ and $p$ into ``large'',
``intermediate''  and ``small'' wave numbers.  We will demonstrate that
the following convergences hold:  there is a sequence of $\rho$
and a subsequence of $\vep$ such that
\begin{align} \label{lc}
&\lim_{\rho \to 0} \lim_{\veps \to 0} \Iouter =\int_{\R^3 \times
\T_*} \D \mu_t(x ,k)\, \overline{b(x,k,\tfrac{k}{|k|})},\\
\label{iwn} &\lim_{\rho \to 0}\lim_{\veps \to 0}
\Ismall^{\mathrm{H}} =
\int_{\R^3 \times S^2} \D \muH_t(x,q)\, \overline{b(x,0,q)},\\
&\lim_{\rho \to 0} \limsup_{\veps \to 0} \left|\Ismall^{\rm wv} -
\mean{a_0,\Wcont^{(1)}[\phi_t]}\right|=0,\label{lwc}\\
&\lim_{\rho \to 0}\limsup_{\veps \to 0} |R|=0.\label{ec}
\end{align}
Clearly, equations (\ref{lc} - \ref{ec}) imply (\ref{eq:entr}).

\subsection*{Large wave numbers.}

We split $\Iouter$ further into two parts using
\begin{align}\label{eq:2ndsplit}
&  1 - \cutoffrac{k_-}{\rho}\cutoffrac{k_+}{\rho}
 = 1- \cutoffrac{k}{\rho}^2 +
\left(\cutoffrac{k}{\rho}-\cutoffrac{k_+}{\rho}\right)\cutoffrac{k}{\rho}
\nonumber \\ & \quad
+\left(\cutoffrac{k}{\rho}-\cutoffrac{k_-}{\rho}\right)\cutoffrac{k_+}{\rho}.
\end{align}
Letting $h_\rho(k) = 1- \cutoffrac{k}{\rho}^2$, which is a smooth function,
the integral then becomes $\Iouter = \Iouter^1 + R_1$, where
\begin{align}%\label{eq:}
&  \Iouter^1 =
\int_{\R^3}\! \D p\, \int_{\Tt}\! \D k\,
\overline{\hat{a}(p,k,\tfrac{k}{\veps})} h_\rho(k) \,
%\nonumber \\ & \qquad \times
\E^{-\I\frac{t}{\veps}\left(\omega(
k +\veps \frac{p}{2}) -\omega( k -\veps \frac{p}{2})\right)}\,
\overline{\hat{\psi}^\vep_0(k -\veps {\tfrac{p}{2}})}\,
\hat{\psi}^\vep_0(k +\veps {\tfrac{p}{2}})\, .
\end{align}
The remainder $R_1$ can be estimated using $|\cutoff|\le 1$ and
\begin{align}%\label{eq:}
\cutoffrac{k_\pm }{\rho}-\cutoffrac{k}{\rho}
= \pm \tfrac{\vep}{2 \rho} \int_0^1\D s\, p \cdot \nabla
\cutoff\Bigl(\tfrac{1}{\rho}(k\pm s \tfrac{\vep}{2}p)\Bigr) ,
\end{align}
which yield the bound, with a universal constant $C$,
\begin{align}%\label{eq:}
  |R_1| \le C \tfrac{\vep}{\rho} \sup_{k,q} \Snorm{a(\cdot,k,q)}{d+2}
  \norm{\hat \psi^\vep_0}^2 \norm{\nabla \cutoff}_\infty.
\end{align}
Therefore, there is a constant $c'$ such that $|R_1|\le c' \vep/\rho$ and thus
$R_1\to 0$ when $\vep\to 0$ for all $\rho$.

We then consider $\Iouter^1$.  The presence of $h_\rho$ guarantees that the
integrand is zero unless $|k|\ge \rho$.  Thus we can
change $\hat a(p,k,k/\vep)$ to $\hat b(p,k,k/|k|)$
in the integrand with an error $R_2$ bounded by
\begin{align}%\label{eq:}
 |R_2| \le C \sup_{k,|q|\ge \rho/\vep}
 \Snorm{a(\cdot,k,q)-b(\cdot,k,\tfrac{q}{|q|})}{d+1}
  \norm{\hat \psi^\vep_0}^2
\end{align}
where $C$ is a universal constant. Therefore, the assumptions imply
that $R_2\to 0$ when $\vep \to 0$ for all $\rho$.  On the other hand,
for $|k|\ge \rho$ and $|p| < \rho/\vep$, inequality (\ref{eq:omkpm})
implies that
\begin{align}%\label{eq:}
\left|\tfrac{1}{\veps}\left(\omega(
k +\veps \tfrac{p}{2}) -\omega( k -\veps \tfrac{p}{2})\right)
-p\cdot \omega(k)\right|
\le C_3 \veps \tfrac{|p|^2}{|k|} \le  C_3 \tfrac{\vep}{\rho} |p|^2.
\end{align}
Therefore, using the estimate $|\E^{\I x}-\E^{\I y}|\le \min(|x-y|,2)$,
valid for all
$x,y\in \R$, we find that we can further change
the $t$-dependent exponential in the integrand to
$\E^{-\I t p \cdot \nabla \omega(k)}$ with a error $R_3$ bounded by
\begin{align}%\label{eq:}
  |R_3| \le C \sup_{k,q} \Snorm{b(\cdot,k,q)}{d+3}
\left(\tfrac{\vep}{\rho} |t|    +
  \int_{|p|\ge \rho/\vep} \D p \sabs{p}^{-d-3}\right)
\end{align}
for some constant $C$.  Therefore, also $\lim_{\vep\to 0}
R_3 =0$ for all $\rho$.
In summary, $\Iouter^1=\Iouter^2 + R_2 + R_3$, where $R_2,R_3$ are
negligible, and
\begin{align}%\label{eq:}
&  \Iouter^2 = \Iouter^2(\vep,\rho) =
 \int_{\R^3}\! \D p\, \int_{\Tt}\! \D k\,
\overline{\hat{b}(p,k,\tfrac{k}{|k|})} h_\rho(k)
\E^{-\I t p \cdot \nabla \omega(k)}\,
%\nonumber \\ & \qquad \times
\overline{\hat{\psi}^\vep_0(k -\veps {\textstyle \frac{p}{2}})}\,
\hat{\psi}^\vep_0(k +\veps {\textstyle \frac{p}{2}})\, .
\end{align}

Let us for a moment consider the lattice Wigner transform
$W_{\text{latt}}^\vep$ of
$\psi_0^\vep$, as  defined in \cite{ls05}.   As pointed out after Definition
\ref{def:Wigners}, then for any
testfunction $f\in \schwartz(\R^3\times \Tt)$, we get an admissible
test-function by the formula $a_f(x,k,q)=f(x,k)$
and then also $\mean{f,W_{\text{latt}}^\vep}= \mean{a_f,\Wep[\psi^\veps_0]}$.
Since $\psi_0^\vep$ is a norm-bounded sequence, the sequence
$W_{\text{latt}}^\vep$ is weak-$*$ bounded, and thus
there is $W_{\text{latt}}^0 \in
\schwartz'(\R^3\times \Tt)$ and a subsequence along which
$W_{\text{latt}}^\vep \wslim
W_{\text{latt}}^0$.

Since the sequence $\psi^{\vep}_0$ is by assumption
also tight on the scale
$\vep^{-1}$, we can then apply Theorems B.4 and B.5 of \cite{ls05} and
conclude that $W_{\text{latt}}^0$
is given by a positive, bounded Radon measure $\mu$
on $\R^3\times \Tt$
such that for all {\em continuous\/} functions $f\in C(\Tt)$ and $p\in \R^3$,
\begin{align}%\label{eq:}
 \lim_{\vep\to 0} \int_{\Tt} \D k \, f(k)
\overline{\hat{\psi}^{\vep}_0(k -\veps {\textstyle \frac{p}{2}})}\,
\hat{\psi}^{\vep}_0(k +\veps {\textstyle \frac{p}{2}})
 = \int_{\R^3\times \Tt} \D \mu(x, k) \, f(k) \E^{-\tpi p\cdot x} .
\end{align}

As $\nabla \omega(k)$ and $k/|k|$ are
continuous apart from $k=0$,
the function
$k\mapsto h_\rho(k) \overline{\hat{b}(p,k,\tfrac{k}{|k|})}
\E^{-\I t p \cdot \nabla \omega(k)}$ is everywhere continuous
for all $\rho>0$ and $p\in \R^3$.  Therefore, by the dominated convergence
theorem, for all $\rho$ we find
\begin{align}%\label{eq:}
& \lim_{\vep\to 0} \Iouter^2 =
 \int_{\R^3}\! \D p\, \int_{\R^3\times \Tt} \D \mu(x, k) \,
\overline{\hat{b}(p,k,\tfrac{k}{|k|})} h_\rho(k) \E^{-\tpi p\cdot
(x+t \nabla \omega(k) (2 \pi)^{-1})} \nonumber \\ & \quad =
\int_{\R^3\times \Tt} \D \mu(x, k) \, h_\rho(k) \overline{b(x+t
\tfrac{1}{2 \pi}\nabla \omega(k),k,\tfrac{k}{|k|})} \, .
\end{align}
When $\rho\to 0$, the integrand approaches pointwise
$\overline{b(x+t \tfrac{1}{2 \pi}\nabla \omega(k),k,\tfrac{k}{|k|})}$
apart from $k=0$, when the limit is $0$.  Therefore, by the dominated
convergence theorem
\begin{align}%\label{eq:}
&\lim_{\rho\to 0} \int_{\R^3\times \Tt} \D\mu(x, k) \, h_\rho(k)
\overline{b(x+t \tfrac{1}{2 \pi}\nabla
\omega(k),k,\tfrac{k}{|k|})} \nonumber \\ & \quad =
\int_{\R^3\times \T_*} \D\mu(x, k) \,
\overline{b(x+t \tfrac{1}{2 \pi}\nabla
\omega(k),k,\tfrac{k}{|k|})}
%\nonumber \\ & \quad
=\int_{\R^3\times \T_*} \D\mu_t(x, k) \,
\overline{b(x,k,\tfrac{k}{|k|})} .
\end{align}
where we have defined the bounded, positive Radon measure $\mu_t$
using $\mu_0=\mu|_{\R^3\times \T_*}$ in the formula
(\ref{eq:defmut}). We have shown that equation (\ref{lc}) holds.

\subsection*{Small wave-numbers and the remainder}

After a change of variables $q =\frac{k}{\veps}$ one obtains that
\begin{align}\label{eq:smallk}
& \Ismall^{\rm wv} =
\int_{\R^3}\! \D p\, \int_{\Tt/\vep}\! \D q\,
\overline{\hat{a}(p,\vep q,q)} \,
\E^{-\I\frac{t}{\veps}\left(\omega(\vep q_+) -\omega(\vep q_-)\right)}\,
\overline{\hat \phi_0(q_-)}\, \hat
\phi_0(q_+)\,\cutoff({\textstyle\frac{\vep q_+}{\rho}})\,
\cutoff({\textstyle\frac{\vep  q_-}{\rho}})\,,
\end{align}
where $q_\pm = q \pm \frac{p}{2}$. We can immediately replace the integration
region for the $q$-integral by $\R^3$.  To see this, note that the
integrand is zero, unless $|q \pm \frac{p}{2}|\le \frac{2 \rho}{\vep}$
for both signs.  Since $2 q =q_+ + q_-$, this can happen only if also
$|q|\le\frac{2 \rho}{\vep}\le \frac{1}{2\vep}$, which implies that the
integrand is zero if $\norm{q}_\infty > \frac{1}{2\vep}$.

However, if $|\vep q|\le 2 \rho$, then
\begin{align}%\label{eq:}
\left| \hat{a}(p,\vep q,q) - \hat{a}(p,0,q) \right|
\le \sup_k\left| \nabla_k\hat{a}(p,k,q)\right| 2 \rho ,
\end{align}
and thus  we can replace in the integrand the function $\hat{a}(p,\vep q,q)$ by
$\hat{a}(p,0,q)$,
with an error $R'_1$ which is bounded by $C \rho$ with a constant $C$
independent of $\vep$ and $\rho$.
Thus we only need to consider the integral
\begin{align}%\label{eq:smallk}
&  \Ismall^1 = \int_{\R^3} \! \D p\,
\int_{\R^3}\! \D q\,
\overline{\hat{a}(p,0,q)} \,
%\nonumber \\ & \qquad \times
\E^{-\I\frac{t}{\veps}\left(\omega(\vep q_+)-\omega(\vep q_-)\right)}\,
\overline{F^{\vep,\rho}(q_-)} F^{\vep,\rho}(q_+) ,
\end{align}
where $F^{\vep,\rho}(q)= \hat \phi_0(q)\cutoff({\textstyle\frac{\vep q}{\rho}})$.
Next we use estimate (\ref{eq:omdiffeq}), which implies
that, if now $|p|\le \frac{1}{2\vep}$, then
\begin{align}\label{eq:omdiffeq3}
\left|\tfrac{1}{\vep}( \omega(\vep q_+) - \omega(\vep q_-)) -
\omega_0(q_+) + \omega_0(q_-) \right|
\le C_4 \veps |p| \, |q| \, .
\end{align}
Following the same argument as earlier, we can then conclude that
the $t$-dependent exponential can be changed to
$\E^{-\I t\left(\omega_0(q_+)-\omega_0(q_-)\right)}$,
with an error $R'_2$ which satisfies the estimate
\begin{align}%\label{eq:}
  |R'_2| \le C \sup_{q} \Snorm{a(\cdot,0,q)}{d+2}
\left( |t| \rho  +
  \int_{|p|\ge (2 \vep)^{-1}} \D p\, \sabs{p}^{-d-2}\right).
\end{align}
Thus $\lim_{\rho\to 0} \limsup_{\vep\to 0} |R'_2| =0$ for all $t$.
Finally, we need to change $F^{\vep,\rho}$ to $\phi_0$, with an error $R'_3$
which can be bounded by $C \norm{F^{\vep,\rho}-\phi_0}$.  Since the bound
goes to zero when $\vep\to 0$ and
\begin{align}%\label{eq:smallk}
&  \Ismall^{\rm wv} = \int_{\R^3} \! \D p\,
\int_{\R^3}\! \D q\, \overline{\hat{a}(p,0,q)} \,
%\nonumber \\ & \qquad \times
\E^{-\I t\left(\omega_0(q_+)-\omega_0(q_-)\right)}\,
\overline{\hat \phi_0(q_-)} \hat \phi_0(q_+) + R'_1+ R'_2+ R'_3,
\end{align}
equation (\ref{lwc}) has been established.

Similar estimates can be employed to demonstrate the vanishing of the
remainder, equation (\ref{ec}).  From the definition of $R$ we get
\begin{align}%\label{eq:}
& R = \int_{\R^3}\! \D p\, \int_{\Tt/\vep}\! \D q\,
\overline{\hat{a}(p,\vep q,q)} \,
\E^{-\I\frac{t}{\veps}\left(\omega(\vep q_+) -\omega(\vep q_-)\right)}\,
\overline{\Bigl(\hat \phi^\vep_0(q_-)-\hat \phi_0(q_-)\Bigr)}\,
\hat \phi_0(q_+)
\nonumber \\ & \qquad \times
\cutoff({\textstyle\frac{\vep q_+}{\rho}})\,
\cutoff({\textstyle\frac{\vep  q_-}{\rho}})
+ \int_{\R^3}\! \D p\, \int_{\Tt/\vep}\! \D q\,
\overline{\hat{a}(p,\vep q,q)} \,
\E^{-\I\frac{t}{\veps}\left(\omega(\vep q_+) -\omega(\vep q_-)\right)}\,
\nonumber \\ & \qquad \times
\overline{\hat\phi_0(q_-)}
\Bigl(\hat \phi^\vep_0(q_+)-\hat \phi_0(q_+)\Bigr)\,
\cutoff({\textstyle\frac{\vep q_+}{\rho}})\,
\cutoff({\textstyle\frac{\vep  q_-}{\rho}}).
\end{align}
We then apply the above estimates to remove the $\vep$-dependence from all
other terms in the integrands, apart from the differences
$\hat{\phi}^\vep_0-\hat \phi_0$.  The error has a bound which
vanishes when $\rho\to 0$.  We are then left with
\begin{align}%\label{eq:}
& \int_{\R^3}\! \D \eta\, \int_{\R^3}\! \D \xi\,
\overline{\hat{a}(\eta-\xi,0,\tfrac{1}{2}(\eta+\xi))} \,
\E^{-\I t\left(\omega_0(\eta) -\omega_0(\xi)\right)}\,
\overline{\Bigl(\hat \phi^\vep_0(\xi)-\hat \phi_0(\xi)\Bigr)}\,
\hat \phi_0(\eta)
\nonumber \\ & \quad
+ \int_{\R^3}\! \D \eta\, \int_{\R^3}\! \D \xi\,
\overline{\hat{a}(\eta-\xi,0,\tfrac{1}{2}(\eta+\xi))} \,
\E^{-\I t\left(\omega_0(\eta) -\omega_0(\xi)\right)}\,
\overline{\hat\phi_0(\xi)}
\Bigl(\hat \phi^\vep_0(\eta)-\hat \phi_0(\eta)\Bigr),
\end{align}
which vanishes as
$\vep\to 0$, since $\phi^\veps_0$ converges weakly to $\phi_0$.  This
establishes (\ref{ec}).

\subsection*{Intermediate wave-numbers}
Changing coordinates $q=\frac{k}{\veps}$ yields that
\begin{align*}
\Ismall^{\mathrm{H}} &=\int_{\R^3}\! \D p\, \int_{\Tt/\vep}\! \D
q\, \overline{\hat{a}(p,\veps q,q)} \,
\E^{-\I\frac{t}{\veps}\left(\omega(\veps q_+) -\omega(\veps
q_-)\right)}\, \overline{\hat{f}^{\vep,\rho}(q_-)} \hat
f^{\vep,\rho}(q_+)\,,
\end{align*}
where $\hat f^{\vep,\rho}(q) = \cutoffrac{\vep q}{\rho}
(\hat\phi^{\vep}_0(q)-\hat\phi_0(q))$.
Let $M>0$ be arbitrary.
We split off the values $|p|> M$ from the integral defining  $\Ismall^3$.
The difference $R'_4=R'_4(\vep,\rho,M)$ can be bounded by
$C\int_{|p|> M} \D p\, \sabs{p}^{-d-1}$ and thus
$\lim_{M\to \infty} \sup_{\rho,\vep} |R'_4| =0$.
We divide the remaining integral over $|p|\le M$ further into two parts using
the identity
\begin{align}\label{eq:sndsplit}
 1 = \ccutoffrac{q_-}{2 M}\ccutoffrac{q_+}{2 M}
 + 1 - \ccutoffrac{q_-}{2 M}\ccutoffrac{q_+}{2 M},
\end{align}
where $\ccutoff=1-\cutoff$. If $|q|\ge 5 M$, then $|q_\pm|\ge 4 M$ and
the second part is zero. It can be checked by inspection that the
sequence $(f^{\vep,\rho})_\vep$ is bounded and tight, and it has a
weak limit zero.
Of these properties only the tightness is non-obvious, but this can also be
easily deduced from the formula
\begin{align}%\label{eq:}
  f^{\vep,\rho}(x) = \rho^3 \vep^{-3/2} \sum_{\gamma\in \Z^3} \psi_0^\vep(\gamma)
  \hat \varphi(\rho(\gamma-\tfrac{1}{\vep}x)) -
 \int_{\R^3}\! \D y\, \phi_0(x + \tfrac{\vep}{\rho} y)  \hat \varphi(y).
\end{align}
Therefore, by Lemma~\ref{th:stcole}, $\lim_{\vep\to
0} \norm{\hat f^{\vep,\rho}}_{L^2(B_{6 M})} =0$ for all $M,\rho$.
This implies that the contribution of the second term, denoted by
$R'_5$, satisfies $\lim_{\vep \to 0} |R'_5| =0$ for all $M,\rho$.

We are thus a left with
\begin{align}%\label{eq:smallk}
& \Ismall^4 = \int_{|p|\le M} \! \D p\,\int_{\R^3}\! \D q\,
\overline{\hat{a}(p,0,q)} \, % \1(|q|\ge M)
%\nonumber \\ & \qquad \times
\E^{-\I t\left(\omega_0(q_+)-\omega_0(q_-)\right)}\,
\overline{\hat g^{\vep,\rho,M}(q_-)} \hat g^{\vep,\rho,M}(q_+) ,
\end{align}
where $\hat g^{\vep,\rho,M}(q)= \ccutoffrac{q}{2 M} \hat f^{\vep,\rho}(q)$
and the integrand can be non-zero only for $|q|\ge M$.
We thus only need to consider $|q|\ge M$ and $|p|\le M$.
First we replace in the integrand
$\hat{a}(p,0,q)$ by $\hat b(p,0,\hat{q})$, $\hat q = \frac{q}{|q|}$, with
an error $R'_6$ which is  bounded by
\begin{align}%\label{eq:}
 |R'_6| \le C \sup_{k,|q|\ge M}
 \Snorm{a(\cdot,k,q)-b(\cdot,k,\tfrac{q}{|q|})}{d+1}
\end{align}
for some constant $C$.  Then we change
$\E^{-\I t\left(\omega_0(q_+)-\omega_0(q_-)\right)}$ to
$\E^{-\I t p \cdot \nabla \omega_0(\hat q)}$ with an error,
$R'_7$ which can be estimated using (\ref{eq:omdiffeq2})
which proves that there is a constant
$C$ such that
\begin{align}%\label{eq:}
 |R'_7| \le C \tfrac{|t|}{M} \sup_{k,q} \Snorm{b(\cdot,k,q)}{d+3}.
\end{align}
Therefore, $\Ismall^4=\Ismall^5+R'_6+R'_7$, where
\begin{align}%\label{eq:smallk}
& \Ismall^5 = \int_{|p|\le M} \! \D p\,\int_{\R^3}\! \D q\,
\overline{\hat{b}(p,0,\hat q)} \, % \1(|q|\ge M)
%\nonumber \\ & \qquad \times
\E^{-\I t p \cdot  \nabla \omega_0(\hat q)}\,
\overline{\hat g^{\vep,\rho,M}(q_-)} \hat g^{\vep,\rho,M}(q_+) .
\end{align}
and $\lim_{M\to \infty} \sup_{\rho,\vep} |R'_6+R'_7|=0$ for all $t$.

Let
\begin{align}%\label{eq:}
b_t(x,\hat q) = b(x+t \tfrac{1}{2\pi}\nabla\omega_0(\hat q),0,\hat q).
\end{align}
Then $b_t\in \schwartz(\R^3\times S^2)$, and it has an extension to a
function $J_t\in \schwartz_6$, i.e., there is $J_t$ such that
$J_t(x,q)=b_t(x,q)$ for all $|q|=1$.  On the other hand, then
\begin{align}%\label{eq:smallk}
&  \Ismall^5 = \int_{|p|\le M} \! \D p\,
\int_{\R^3}\! \D q \,
\overline{\hat{b}_t(p,\hat q)} \,
\overline{\hat g^{\vep,\rho,M}(q_-)} \hat g^{\vep,\rho,M}(q_+) =
\mean{J_t,\Lambda^{\vep,\rho,M}}
\end{align}
where $\Lambda^{\vep,\rho,M}\in \schwartz'_6$ denotes the distribution
\begin{align}%\label{eq:smallk}
& J\mapsto \mean{J,\Lambda^{\vep,\rho,M}}=\int_{|p|\le M} \! \D p\,
\int_{\R^3}\! \D q \,
\overline{\hat{J}(p,\hat q)} \,
\overline{\hat g^{\vep,\rho,M}(q_-)} \hat g^{\vep,\rho,M}(q_+) .
\end{align}

Clearly, each $\Lambda^{\vep,\rho,M}$ has support in $\R^3\times S^2$, and
there is a constant $C$ such that for all $J,\vep,\rho,M$
\begin{align}%\label{eq:}
|\mean{J,\Lambda^{\vep,\rho,M}}|\le C \norm{\hat g^{\vep,\rho,M}}^2 \Snorm{J}{d+1}.
\end{align}
However, since we have
$\norm{\hat g^{\vep,\rho,M}}\le \norm{\hat f^{\vep,\rho}}$,
where $\norm{\hat f^{\vep,\rho}}$ is
bounded in $\vep$, Banach-Alaoglu theorem implies that   the family
$(\Lambda^{\vep,\rho,M})_{\vep,\rho,M}$ belongs to a
weak-$*$ sequentially compact set.  Therefore, for every $\rho,M$
there is a subsequence of $(\vep)$ and $\Lambda^{\rho,M}$ such that
$\Lambda^{\vep,\rho,M}\overset{*}{\wlim} \Lambda^{\rho,M}$ along this
subsequence.  In addition, for every $\rho$ there is $\Lambda^{\rho}$
and a sequence of integers $M$ such that
$\Lambda^{\rho,M}\overset{*}{\wlim} \Lambda^{\rho}$ along this
sequence. Finally, there is $\Lambda \in \schwartz'_6$ and a sequence
of integers $N\ge 4$ such that for $\rho =\frac{1}{N}$,
$\Lambda^{\rho}\overset{*}{\wlim} \Lambda$.

All of the above distributions clearly must have support on $\R^3\times S^2$.
We will soon prove that, in addition, for all $f\in\schwartz_6$ and $\rho$
\begin{align}\label{eq:Lambdapos}
 \mean{|f|^2,\Lambda^\rho}\ge 0 .
\end{align}
This implies then that also $\mean{|f|^2,\Lambda}\ge 0$.
Therefore, by the Bochner-Schwartz theorem, there is a positive
Radon measure $\muH$ on $\R^6$ such that for all testfunctions
$J$, $\mean{J,\Lambda} = \int \muH(\D x,\D k) \, J(x,k)$. Since
also $\muH$ must have support on $\R^3\times S^2$, we can thus
identify it with a positive Radon measure $\muH_0$ on $\R^3\times
S^2$.  By considering testfunctions $J(x,k)=\E^{-\delta^2 x^2}$
in the limit $\delta\to 0$, it is also clear that $\muH_0$ must be
bounded.  We then define the positive, bounded Radon measures
$\muH_t$, $t\in \R$, by the formula (\ref{eq:defmuht}). It follows
from the construction of $\muH_t$ that equation (\ref{iwn}) holds along the
above sequences $\rho,\vep$.

The main missing ingredient is provided by the following Lemma
\begin{lemma}\label{th:fsfest}
For $p,q\in \R^3$, let us denote $q_\pm = q \pm \frac{p}{2}$, $\hat
q_\pm = q_\pm/|q_\pm|$, and $\hat q = q/|q|$. There is a constant $C$
such that for all $q,q'\in \R^3$, and $f\in\schwartz_6$
  \begin{align}\label{eq:fsfgenb}
    \left| \int_{\R^3} \D x\, \E^{-\tpi p\cdot x}
      \overline{f(x,q')} f(x,q) \right|\le
    C \sabs{p}^{-d-1} \Snorm{f}{d+1}^2.
  \end{align}
If, in addition, $|q|\ge M$ and $|p|\le M$, then also
  \begin{align}\label{eq:fsfdiffb}
    \left| \mathcal{F}_{x\to p}(|f|^2)(p,\hat{q})  -
  \int_{\R^3} \D x\, \E^{-\tpi p\cdot x}
  \overline{f(x,\hat q_+)} f(x,\hat q_-) \right|\le
   \tfrac{1}{M} C \sabs{p}^{-d-1} \Snorm{f}{d+2}^2.
  \end{align}
\end{lemma}
Before proving the lemma we demonstrate that it implies
inequality~(\ref{eq:Lambdapos}). Let then $f\in\schwartz_6$ be
arbitrary. Define $q_\pm,\hat q_\pm$ as in Lemma~\ref{th:fsfest},
except that let here also $\hat 0 = 0$, and let
\begin{align}\label{eq:deffinalI}
& I^{\vep,\rho,M,f} = \int_{\R^3}\! \D p\,
\int_{\R^3}\! \D q \,
\Bigl[  \int_{\R^3} \D x \, \E^{\tpi p\cdot x}
 f(x,\hat q_+)  \overline{f(x,\hat q_-)} \Bigr]\,
\overline{\hat g^{\vep,\rho,M}(q_-)} \hat g^{\vep,\rho,M}(q_+) .
\end{align}
Note that this integral is well-defined by (\ref{eq:fsfgenb}).
Then, by estimate (\ref{eq:fsfdiffb}) and uniform boundedness of
$g^{\vep,\rho,M}$, there is a constant $c$ such that
\begin{align}\label{eq:f2Iest}
& \left|\mean{|f|^2,\Lambda^{\vep,\rho,M}}-I^{\vep,\rho,M,f}\right|
%\nonumber \\ & \quad
\le c \Snorm{f}{d+2}^2 \left(\int_{|p|\ge M} \D p\, \sabs{p}^{-d-1}
 +   \tfrac{1}{M}\right).
\end{align}
On the other hand, always $I^{\vep,\rho,M,f}\ge 0$. To see this, consider
first the case when $g^{\vep,\rho,R}\in \schwartz$.  Then
$\hat g^{\vep,\rho,R}\in \schwartz$, and by
changing variables from $(q,p)$
to $(q_+,q_-)$ in (\ref{eq:deffinalI}), and then using Fubini's theorem to
reorder the integrals, we find that
\begin{align}\label{eq:Iispos}
& I^{\vep,\rho,M,f} = \int_{\R^3} \D x  |G(x)|^2,\qquad \text{with }
 G(x) = \int_{\R^3} \D q\, \E^{\tpi q\cdot x} f(x,\hat q)
 \hat g^{\vep,\rho,R}(q) \in L^2.
\end{align}
Since $I^{\vep,\rho,M,f}$ depends $L^2$-continuously on $g^{\vep,\rho,R}$
this implies that also for general $g^{\vep,\rho,R}\in L^2$
we have $I^{\vep,\rho,M,f}\ge 0$.  Since the right hand side of
(\ref{eq:f2Iest}) vanishes if first $\vep\to 0$ and then $M\to
\infty$, we must thus also have $\mean{|f|^2,\Lambda^\rho}\ge 0$.
This proves (\ref{eq:Lambdapos}).

The only remaining task is to prove Lemma~\ref{th:fsfest}.
Consider first (\ref{eq:fsfgenb}).  If $|p|\le 1$, we have
trivially a bound
  \begin{align}%\label{eq:fsfgenb}
   \int_{\R^3} \D x\, |f(x,q')|\, |f(x,q)|\le
    \Snorm{f}{d+1}^2  \int_{\R^3} \D x\, \sabs{x}^{-2 d-2} \le c
    \Snorm{f}{d+1}^2.
  \end{align}
If $|p|\ge 1$, we perform $N$ partial integrations in the
direction of $p$, that is in the direction $\hat p =
\frac{p}{|p|}$, yielding
  \begin{align}%\label{eq:fsfgenb}
 \int_{\R^3} \D x\, \E^{-\tpi p\cdot x}
      \overline{f(x,q')} f(x,q) =
\tfrac{1}{(\tpi |p|)^N} \int_{\R^3} \D x \E^{-\tpi p\cdot x}
     (\hat p\cdot \nabla)^N \left(\overline{f(x,q')} f(x,q) \right).
  \end{align}
By the Leibniz rule,
\begin{align}%\label{eq:}
     (\hat p\cdot \nabla)^N \left(\overline{f(x,q')} f(x,q) \right)
 = \sum_{n=0}^N \binom{N}{n}
(\hat p\cdot \nabla)^n \overline{f(x,q')} (\hat p\cdot
\nabla)^{N-n} f(x,q)
\end{align}
which is bounded by $c'_N \sabs{x}^{-N} \Snorm{f}{N}^2$.  Choosing
$N=d+1$ then yields (\ref{eq:fsfgenb}) for some constant.
Adjusting the constant $C$ so that the bound is true also for
$|p|\le 1$ proves that (\ref{eq:fsfgenb}) is valid.

To prove (\ref{eq:fsfdiffb}), consider $q,p$ as required in the
Lemma.  Let $q(s)= q + \frac{s}{2}p$, when $|q(s)|\ge |q|/2>0$ for
all $|s|\le 1$, and $\hat q(s)$ is thus well-defined and smooth.
Therefore, for any $g\in\schwartz_6$ and $s_0\in [-1,1]$,
\begin{align}%\label{eq:}
& g(x,\hat q(s_0))-g(x,\hat q) = \int_0^{s_0} \D s\, \tfrac{\D}{\D
s}
  g(x,\hat q(s))
\nonumber \\ & \quad
 = \int_0^{s_0} \D s\,
 \left(\tfrac{1}{2 |q(s)|}p- \tfrac{p\cdot \hat q(s)}{2 |q(s)|} \hat q(s)
  \right)\cdot\left. \nabla_q g(x,q)\right|_{q=\hat q(s)}
\end{align}
implying
\begin{align}%\label{eq:}
|g(x,\hat q_\pm)-g(x,\hat q)| \le 2\tfrac{|p|}{|q|} \sup_{|q|=1}
 |\nabla_q g(x,q)|.
\end{align}
Also for all $g_1,g_2\in \schwartz_6$,
\begin{align}%\label{eq:}
& \overline{g_1(x,\hat q)} g_2(x,\hat q) - \overline{g_1(x,\hat
q_-)} g_2(x,\hat q_+) \nonumber \\ & \quad =
\overline{\left(g_1(x,\hat q)-g_1(x,\hat q_-)\right)} g_2(x,\hat
q)
  +  \overline{g_1(x,\hat q_-)} \left(g_2(x,\hat q)-g_2(x,\hat q_+)\right).
\end{align}
Following the steps made in the first part of the proof, and
replacing the earlier estimates with the above more accurate ones
when necessary, we can conclude that the constant $C$ can be
adjusted so that for these values of $q,p$ also
(\ref{eq:fsfdiffb}) holds.
This completes the proof of equation (\ref{iwn}).

\subsection*{Energy equality}
The energy equality (\ref{eq:eneq})
follows by considering a sequence of testfunctions
$a^\delta=\E^{-\delta^2 x^2}$ and taking $\delta \to 0$.  To see
this, first note that the right hand side of (\ref{eq:eneq}) is
clearly independent of $t$, and thus it is enough to consider
$t=0$. Thanks to equation (\ref{eq:enrel}) and to the tightness of
the sequence $\psi_0^\veps$ we obtain for this particular
testfunction that
\begin{align*}
 \lim_{\delta \to 0}
\lim_{\veps \to 0} \mean{a^\delta,\Wep[\psi^\veps_0]} = \lim_{\delta\to 0}\lim_{\veps\to 0} \mean{a^\delta,e^\veps}=  \lim_{\vep \to 0} \|\psi^\veps_0\|^2
\end{align*}
Equation (\ref{eq:entr}) implies that
\begin{align*}
\lim_{\veps \to 0}\mean{a^\delta,\Wep[\psi^\veps_0]} =
\int_{\R^3}\dd x\, a^\delta |\phi_0(x)|^2 + \int_{\R^3}
\int_{\Tt}a^\delta(x)\,\dd \mu_0(x,k)+\int_{\R^3}
\int_{S^2}a^\delta(x)\,\dd \muH_0(x,k).
\end{align*}
Sending $\delta$ to 0 yields that
$$ \lim_{\vep \to 0} \|\hat \psi^\vep_0\|^2 =
\|\phi_0\|_{L^2(\R^3)}^2 + \mu_0(\R^3\times \Tt) +
\muH_0(\R^3\times S^2),$$
and the energy equality has been
established.  This finishes the proof of Theorem~\ref{th:mainres}.
%\end{proofof}

\section{Proof of Corollary \ref{th:maincorr}}\label{sec:corrproof}

%\begin{proof}
Let $I_0$ denote the original sequence of $\vep$, and consider an arbitrary
subsequence $I$ of $I_0$.  Since $(\psi^\vep_0)_{\vep\in I}$
then also satisfies
Assumption~\ref{th:mainassump}, we can conclude from
Theorem~\ref{th:mainres} that for every $I$ there is a subsequence $I'$
such that (\ref{eq:entr}) holds for all $t$ with the initial conditions
given by some triplet
$(\mu_{I},\muH_{I},\phi_{I})$.  From the construction of the subsequence
in the proof of Theorem~\ref{th:mainres}, we know that
$\phi_I$ can be chosen as the weak limit of $\phi^\vep_0$ along the
subsequence $I'$.  The first assumption thus implies that
we can always choose $\phi_I=\phi_0$.  Let us also denote $\mu_0=\mu_{I_0}$
and $\muH_0=\muH_{I_0}$, and to prove the stated uniqueness, we will prove
that $\mu_I =\mu_{0}$ and $\muH_I=\muH_{0}$ for all $I$.

Consider first an arbitrary $\tilde a\in C^{(\infty)}_c(\R^3\times \Tt_*)$.
Let $a(x,k,q)=\tilde a(x,k)$ for $k\ne 0$ and define $a(x,0,q)=0$.
Then $a$ is an admissible test-function with $a_0=0=b(k,0,q)$ and
for any subsequence $I$ we thus obtain,
using the second assumption
\begin{align}%\label{eq:}
& \int_{\R^3 \times \Tt_*} \dd \mu_{I}(x, k)\, \overline{\tilde a(x,k)}
 = \lim_{\vep \in I} \mean{a,\Wep[\psi^\veps_0]}
 =  \int_{\R^3 \times \Tt_*} \dd \mu_{0}(x, k)\, \overline{\tilde a(x,k)}.
\end{align}
Such $\tilde a$ are dense in $C_c(\R^3\times \Tt_*)$, and
thus $\mu_{I}=\mu_{0}$ on $\R^3\times \Tt_*$.

Consider then an arbitrary $b\in C^{(\infty)}_c(\R\times S^2)$.
Let $\varphi$ be a smooth cutoff function as in Definition
\ref{def:cutoff}, and define
$a(x,k,q)=b(x,q/|q|)\,(1-\varphi(2 q))$.  Then $a$ is an admissible
testfunction and indeed
$\lim_{R\to \infty} a(x,k,R q)=b(x,q)$ for all $|q|=1$.
By the already proven results
we then find that for any subsequence $I$
\begin{align} % \label{eq:entr}
& \int_{\R^3 \times S^2} \dd \muH_I(x,q)\,
\overline{b(x,q)}\, =
\lim_{\veps\to 0} \mean{a,\Wep[\psi^\veps_0]} -
\int_{\R^3 \times \T_*} \dd \mu_0(x, k)\,
\overline{b(x,\tfrac{k}{|k|})}\,
\nonumber \\ & \qquad
 - \mean{a_0,\Wcont^{(1)}[\phi_0]}
= \int_{\R^3 \times S^2} \dd \muH_0(x,q)\,
\overline{b(x,q)}\, .
\end{align}
Therefore, also $\muH_I=\muH_{0}$, which concludes the uniqueness part of
the proof of the Corollary.

Finally, define for $t\ne 0$ the triplet
$(\mu_{t},\muH_{t},\phi_{t})$ using $(\mu_0,\muH_0,\phi_0)$
as the initial data.  By the above proved uniqueness,
for any subsequence $I$,
(\ref{eq:entr}) holds along the subsubsequence $I'$.
As the right hand side is thus independent of
$I$, this proves that the limit also holds along the original sequence
$I_0$.
This completes the proof of the Corollary.
%\end{proof}

\appendix

%\section{Estimates for the dispersion relation}
\section{Technicalities}

The proof of Theorem~\ref{th:mainres} relies on two simple lemmas which are
provided here. The first lemma summarizes several properties of
regular acoustic dispersion relations.
\begin{lemma}\label{th:dispprop}
Let $\omega$ be a regular acoustic dispersion relation and recall the
definitions of $\lambda$, $A_0$ and $\omega_0$.
Then the following assertions are true.
\begin{enumerate}
\item\label{it:omreg}
$\omega\in C^{(3)}(\T\backslash \left\{0\right\},\R)$.
\item\label{it:alphaorig} $\nabla \lambda(0) =0$ and $A_0 > 0$.
\item\label{it:firstb} There are constants $C_1,C_2>0$ such that for all
$\mod{k}_\infty\le \frac{3}{4}$,
\begin{align}\label{eq:firstb}
  \omega(k) \ge C_1 |k|, \qquad |\nabla \lambda (k)|\le C_2 |k|\, .
\end{align}
In addition, $\norm{\nabla \omega}_{\infty} <\infty$.
\item\label{it:omdiff} There is $C_3$ such that,
if $\veps>0$, $p\in \R^3$, and $\mod{k}_\infty\le \tfrac{1}{2}$, with
$|k|> \veps |p|$, then
%for $k_\pm = k \pm \frac{1}{2}\veps p$,
\begin{align}\label{eq:omkpm}
 | \omega(k + \tfrac{1}{2}\veps p) - \omega(k - \tfrac{1}{2}\veps p)
 - \veps p \cdot \nabla \omega(k) | \le C_3 \veps^2 \tfrac{|p|^2}{|k|}.
\end{align}
\item\label{it:smallom}
There is $C_4$ such that, if $\veps> 0$ and $p,q\in\R^3$
with $|p|,\norm{q}_\infty\le \frac{1}{2}\vep^{-1}$, then
for $q_\pm = q \pm \frac{1}{2} p$,
\begin{align}\label{eq:omdiffeq}
| \omega(\vep q_+) - \omega(\vep q_-) -
\omega_0(\vep q_+) + \omega_0(\vep q_-) |
\le C_4 \veps^2 |p| \, |q| \, .
\end{align}
\item\label{it:largeom0}
There is $C_5$ such that, if $p,q\in\R^3$ with $q\ne 0$ and
$|p|\le|q|$, then for $q_\pm = q \pm \frac{1}{2} p$,
\begin{align}\label{eq:omdiffeq2}
|\omega_0(q_+) - \omega_0(q_-)
-p\cdot\nabla \omega_0(\tfrac{q}{|q|})|
\le C_5 \tfrac{|p|^2}{|q|}.
\end{align}
\end{enumerate}
\end{lemma}
\begin{proof}
The first item is obvious, and the second one follows from the
assumptions, since $0$ is a minimum. The second inequality in
%item \ref{it:firstb}
(\ref{eq:firstb})
follows by using item \ref{it:alphaorig}
and $\norm{D^2 \lambda}_\infty < \infty$, when by Taylor expansion
$|\nabla \lambda (k)|\le C_2 |k|$ for all $k\in \R$.
To prove the first inequality, we first note that, by continuity, also
$\norm{D^3 \lambda}_\infty < \infty$.  Thus there is $c'$ such that for
all $k\in\R^3$
\begin{align}\label{eq:gdiffbound}
 | \lambda(k) - \lambda_0(k) | \le c' |k|^3 ,
\end{align}
where $\lambda_0(k) = \frac{1}{2} k\cdot A_0 k$.  Since $A_0>0$, there is $c>0$
such that $\lambda_0(k)\ge c^2 k^2$ for all $k$.
Thus there is $\delta'>0$
such that for all $|k|<\delta'$, we have $|1-\lambda(k)/\lambda_0(k)|\le
\frac{3}{4}$, and for these $k$ therefore also
$\omega(k) = \sqrt{\lambda(k)-\lambda_0(k)+\lambda_0(k)}
\ge \frac{1}{2} \sqrt{\lambda_0(k)}\ge \frac{c}{2} |k|$.  Since
$\lambda$ has no zeroes in the complement set,  the
constant can then be adjusted so that (\ref{eq:firstb}) holds for all
$k$ with $\mod{k}_\infty\le \frac{3}{4}$.

We still need to prove the third
property, boundedness of $\nabla\omega$.  For later use, let us,
more generally, consider a non-negative $f\in C^{(2)}(\R^3)$,
$q\in \R^3$, and $k$ such that $f(k)\ne 0$.  Then we have
\begin{align}%\label{eq:}
 (q\cdot \nabla) \sqrt{f(k)}
& =   \frac{1}{2\sqrt{f(k)}} q\cdot \nabla f(k), \\
 (q\cdot \nabla)^2 \sqrt{f(k)}
& = -\frac{1}{4 f(k)^{3/2}} (q\cdot \nabla f(k))^2
 +\frac{1}{2\sqrt{f(k)}} (q\cdot \nabla)^2 f(k).
\end{align}
This implies that there is
a constant $C$ such that for all $q\in \R^3$ and $k\ne 0$, with
$\mod{k}_\infty \le \frac{3}{4}$,
\begin{align}%\label{eq:}
  |q\cdot \nabla \omega(k)|, |q\cdot \nabla \omega_0(k)|
& \le C  |q|\, , \\
 |(q\cdot \nabla)^2 \omega(k)|, |(q\cdot \nabla)^2 \omega_0(k)|
& \le C  \frac{|q|^2}{|k|}\, .
\end{align}
In particular, by periodicity therefore
$\norm{\nabla\omega}_\infty<\infty$.

To prove item \ref{it:omdiff}, consider $k,p,\veps$ as in the claim, and
define a function
$f(s)=\omega(k_+)-\omega(k_-)$ with $k_{\pm} = k\pm s \frac{1}{2}p$
and $s\in [-\veps,\veps]$.
Then $\mod{k_{\pm}}_\infty\le \frac{3}{2} \mod{k}_\infty$ and
$|k_{\pm}|\ge |k|-\veps \frac{1}{2}p \ge \frac{1}{2}|k|>0$, and thus $f$
belongs to $C^{(3)}$.  In particular, $f(0)=0$,
$f'(0)=p\cdot \nabla\omega(k)$, and
\begin{align}
& f''(s) = \frac{1}{2} \Bigl[
\frac{1}{\omega(k_+)} \left(\frac{p}{2}\cdot \nabla\right)^2\! \lambda(k_+)
-\frac{1}{\omega(k_-)} \left(\frac{p}{2}\cdot \nabla\right)^2\! \lambda(k_-)
\nonumber \\ & \qquad
-\frac{1}{2 \omega(k_+)^3} \left(\frac{p}{2}\cdot \nabla \lambda(k_+)\right)^2
+\frac{1}{2 \omega(k_-)^3} \left(\frac{p}{2}\cdot \nabla \lambda(k_-)\right)^2
 \Bigr] .
\end{align}
Using item \ref{it:firstb}, we find that there is $C$ such that
this is uniformly bounded by $C|p|^2/|k|$.  Then a Taylor expansion
at the origin proves item \ref{it:omdiff}.

Since for all $k\ne 0$,
\begin{align}
 \omega(k) - \omega_0(k) = \sqrt{\lambda(k)}-\sqrt{\lambda_0(k)} =
\frac{\lambda(k)-\lambda_0(k)}{\sqrt{\lambda(k)}+\sqrt{\lambda_0(k)}}\,,
\end{align}
and then also
\begin{align}%\label{eq:}
& q\cdot \nabla (  \omega(k) - \omega_0(k) )
\nonumber \\ & \quad
= \frac{q\cdot \nabla \lambda(k)-q\cdot \nabla\lambda_0(k)}{
  \omega(k)+\omega_0(k)} -
(\lambda(k)-\lambda_0(k))
\frac{q\cdot \nabla \omega(k) + q\cdot \nabla \omega_0(k)}{
  (\omega(k)+\omega_0(k))^2}  \, .
\end{align}
By Taylor expansion at the origin, we find that
there is $C'$ such that for all $q\in \R^3$,
\begin{align}%\label{eq:}
& |\lambda(k)-\lambda_0(k)| \le C' |k|^3, \qquad
|q\cdot \nabla \lambda(k)- q\cdot \nabla \lambda_0(k)| \le C' |k|^2 |q|, \\
& |(q\cdot \nabla)^2 \lambda(k)- (q\cdot \nabla)^2 \lambda_0(k)| \le
C' |k|\, |q|^2 \, .
\end{align}
Thus there is also $C$ such that for all $q\in \R^3$, and
$\mod{k}_\infty \le \frac{3}{4}$, with $k\ne 0$,
\begin{align}\label{eq:nablaom0}
& |q\cdot \nabla (  \omega(k) - \omega_0(k) ) |\le
C  |k|\, |q|\, ,\qquad
 |(q\cdot \nabla)^2 (  \omega(k) - \omega_0(k) ) |\le
C  |q|^2 .
\end{align}

Let us then consider $q,p,\vep$ satisfying the assumptions made in the
final item.  Since then $\norm{\vep q_\pm}_\infty \le \frac{3}{4}$, we can
apply the previous estimates.
If $p=0$, then $q_-=q_+$ and
the bound in (\ref{eq:omdiffeq}) is trivially valid for any $C_4$.
Consider thus $p\ne 0$, and assume first that $p$ is not proportional to
$q$.  Since then the line segment $[0,1]\ni s\mapsto \vep q + s \vep p/2$
does not pass through the origin,
the function
$s\mapsto \omega(\vep q + s \vep p/2)-\omega_0(\vep q + s \vep p/2)$ is in
$C^{(3)}([0,1])$.
We make a Taylor expansion of this function at $s=0$, yielding
\begin{align}%\label{eq:}
\omega(\vep q + \vep \tfrac{p}{2})-\omega_0(\vep q + \vep\tfrac{p}{2})
= \omega(\vep q)-\omega_0(\vep q) +
\vep \tfrac{p}{2} \cdot (\nabla \omega(\vep q) - \nabla \omega_0(\vep q))
+ R\, .
\end{align}
Here (\ref{eq:nablaom0}) implies $|R|\le C \vep^2 |p|^2$, since
\begin{align}%\label{eq:}
R = \int_0^1 \!\D s\, (1-s)
 \Bigl(\vep \tfrac{p}{2}\cdot \nabla\Bigr)^2
 (\omega(\vep q + s \vep \tfrac{p}{2})-
 \omega_0(\vep q + s\vep \tfrac{p}{2})) \, .
\end{align}
This proves that (\ref{eq:omdiffeq}) holds in this case for some constant
$C_4$.   In final remaining case $p\propto q$, we choose a direction $u$
orthogonal to $q$, and use the previous estimate with $p+\delta u$ instead
of $p$ for an arbitrary $0<\delta\le 1$.  Since the left hand side
of (\ref{eq:omdiffeq}) is continuous in $\delta$, the bound must then hold
also for $\delta=0$, proving the validity of the estimate also in this case.

To prove (\ref{eq:omdiffeq2}), we use the fact that by assumption
$|q_\pm|\ge \frac{1}{2}|q|>0$, and thus denoting $\hat q= q/|q|$, we get
\begin{align}%\label{eq:}
\omega_0(q_+) - \omega_0(q_-) = |q|
\left( \omega_0(\hat q +\frac{p}{2|q|}) -
  \omega_0(\hat q - \frac{p}{2|q|})\right)
= p \cdot\nabla \omega_0(\hat q) + R
\end{align}
where $|R|\le C |p|^2 /|q|$.  We have thus completed the
proof of the Lemma.
\end{proof}
The second lemma recalls a well-known fact in Fourier analysis: the
Fourier transform of a bounded and tight sequence of $L^2$ functions
converges strongly on compact sets. For the convenience of the reader
we give a proof.
\begin{lemma} \label{th:stcole}
Let
$f^\veps \in L^2(\R^d)$ be a bounded and tight sequence of
functions such that $f^\veps \rightharpoonup 0$ as $\veps \to 0$.
Then
\begin{equation}
\label{eq:strong} \lim_{\veps\to 0} \|{\wh f}^\veps\|_{L^2(\Omega)}
=0.
\end{equation}
for every $\Omega\subset \R^d$ with a finite measure.
\end{lemma}
\begin{proof}
Choose an arbitrary number $\beta>0$. By tightness
of $f^\veps$ there exists a number $R_\beta>0$ such that
\begin{equation} \label{eq:tightprop}
\limsup_{\veps\to 0} \int_{|x| \geq R_\beta}\dd x\,
|f^\veps(x)|^2\leq \beta^2.
\end{equation}
Define now
\begin{align}%\label{eq:}
f^{\veps,\beta}(x) = \begin{cases} f^\veps(x) & \text{if
  } |x| \leq R_\beta,\\
0 & \text{else.}
\end{cases}
\end{align}
By the boundedness of $f^{\veps,\beta}$ in $L^2(\R^d)$ there exists
$g^{\beta}$ and a subsequence $(\vep')$ such that $f^{\veps',\beta}
\wlim g^\beta$ in $L^2(\R^d)$ as $\veps' \to 0$. Estimate
(\ref{eq:tightprop}) implies that $\limsup_{\veps \to 0} \|
f^\veps-f^{\veps,\beta}\|^2_{L^2(\R^3)} \leq \beta^2$.
Since also $f^\veps \wlim 0$, we have
\begin{align}%\label{eq:}
\norm{g^\beta}^2 = \lim_{\veps'}
|\mean{g^\beta,f^{\veps',\beta}}|
\le \norm{g^\beta}\limsup_{\veps \to 0}\norm{f^\veps-f^{\veps,\beta}}
\end{align}
and, therefore, $\|g^\beta\|_{L^2(\R^d)} \leq \beta$.

Since $g^\beta$ has support in a ball of radius $R_\beta$,
also $g_1^\beta(x) = -\tpi x g^\beta(x) \in L^2(\R^d)$.  Therefore,
$\hat g^\beta \in H^1(\R^d)$, and $\nabla \hat g^\beta = \hat g_1^\beta$.
Similarly, also $\hat f^{\veps,\beta}\in H^1(\R^d)$ and it is
straightforward to check that $\nabla \hat f^{\veps',\beta} \wlim
\nabla \hat g^\beta$ in $L^2$.
Since $\Omega$ is assumed to be a set of finite measure,
we have then
$\lim_{\veps' \to 0} \|\wh f^{\veps',\beta} - \wh
f^\beta\|_{L^2(\Omega)}=0$ (for a proof, see for instance Theorem 8.6.\ in
\cite{ll:analysis}).
This result yields the following
estimate for the original sequence $\wh f^\veps$:
\begin{align*}
& \limsup_{\veps' \to 0} \|\wh f^{\veps'}\|_{L^2(\Omega)}
\nonumber \\ & \quad
\leq
\limsup_{\veps' \to 0} \|\wh f^{\veps'}-\wh
f^{\veps',\beta}\|_{L^2(\Omega)} + \limsup_{\veps' \to 0} \|\wh
f^{\veps',\beta}-\wh f^{\beta}\|_{L^2(\Omega)} +
\| \wh f^\beta\|_{L^2(\Omega)} \le 2 \beta.
\end{align*}
Since $\beta$ can be arbitrarily small, we obtain that there must be a
subsequence $\vep''$ such that
$\lim_{\veps'' \to 0} \|\wh f^{\veps''}\|_{L^2(\Omega)} =0$.

Since the assumptions on the sequence
$f^\vep$ are preserved for subsequences, we can consider an arbitrary
subsequence and apply the above result to it.  Then we can conclude that
for every subsequence there is a subsubsequence along which
(\ref{eq:strong}) holds.  This implies
that the limit (\ref{eq:strong}) actually holds also along the original
sequence.
\end{proof}

% \newcommand{\utildir}[1]{../../../texstuff/#1}
% \bibliographystyle{\utildir{abunst_titles}}
% \bibliography{\utildir{myabbr},\utildir{mrabbrev},\utildir{allrefs}}
% \end{document}

\end{document}